\documentclass[traditabstract]{aa}
\usepackage{epsfig}    
\usepackage{lscape}
\usepackage{natbib}
\bibpunct{(}{)}{;}{a}{}{,}    
\usepackage{graphics}
\usepackage{graphicx,graphics}
\usepackage{color} 
\usepackage{times}
\usepackage{amsmath}
\usepackage{amssymb}
\begin{document}

\title {Validity of abundances derived from spaxel spectra of the MaNGA survey} 
      
\author {L.~S.~Pilyugin\inst{\ref{IC},\ref{MAO},\ref{ARI}} \and 
        E.~K.~Grebel\inst{\ref{ARI}} \and
        I.~A.~Zinchenko\inst{\ref{MAO},\ref{ARI}} \and
        Y.~A.~Nefedyev\inst{\ref{KGU}} \and
        V.~M.~Shulga\inst{\ref{IC},\ref{RIAN},\ref{CP}} \and 
        Han Wei\inst{\ref{IC},\ref{CP}} \and 
        P.P.~Berczik\inst{\ref{IC},\ref{MAO},\ref{ARI}} }
\institute{The International Center of Future Science of the Jilin University, 2699 Qianjin St., 130012, Changchun City, China \label{IC} \and 
Main Astronomical Observatory, National Academy of Sciences of Ukraine, 27 Akademika Zabolotnoho St, 03680, Kiev, Ukraine \label{MAO} \and
Astronomisches Rechen-Institut, Zentrum f\"{u}r Astronomie der Universit\"{a}t Heidelberg, M\"{o}nchhofstr.\ 12--14, 69120 Heidelberg, Germany \label{ARI} \and
Kazan Federal University, 18 Kremlyovskaya St., 420008, Kazan, Russian Federation \label{KGU} \and
Institut of Radio Astronomy of National Academy of Sciences of Ukraine, 4 Mystetstv str., 61002 Kharkov, Ukraine  \label{RIAN} \and 
College of Physics, The Jilin University, 2699 Qianjin St., 130012, Changchun, China \label{CP} }

\abstract{
We measured the emission lines in the spaxel spectra of MaNGA (Mapping
Nearby Galaxies at Apache Point Observatory) galaxies in order to
determine the abundance distributions therein.  It has been
suggested that the strength of the low-ionization lines, R$_{2}$,
N$_{2}$, and S$_{2,}$ may be increased (relative to Balmer lines) in
(some) spaxel spectra of the MaNGA survey due to a contribution of the
radiation of the diffuse ionized gas.  Consequently, the abundances
derived from the spaxel spectra through strong-line methods may suffer
from large errors.  We examined this expectation by comparing the
behaviour of the line intensities and the abundances estimated through
different calibrations for slit spectra of H\,{\sc ii} regions in
nearby galaxies, for fibre spectra from the Sloan Digital Sky Survey
(SDSS), and for spaxel spectra of the MaNGA survey.  We found that the
S$_{2}$ strength is increased significantly in the fibre and spaxel
spectra. The mean enhancement changes with metallicity and can be as
large as a factor of $\sim 2$.  The mean distortion of R$_{2}$ and
N$_{2}$ is less than a factor of $\sim 1.3$.  This suggests that
Kaufmann et al.'s demarcation line between active galactic nuclei (AGNs) and H\,{\sc ii}
regions in the Baldwin, Phillips, \& Terlevich (BPT) diagram is a
useful criterion to reject spectra with significantly distorted
strengths of the N$_{2}$ and R$_{2}$ lines.  We find that the
three-dimensional $R$ calibration, which uses the N$_{2}$ and R$_{2}$
lines, produces reliable abundances in the MaNGA galaxies.  The
one-dimensional N2 calibration produces either reliable or wrong
abundances depending on whether excitation and N/O abundance ratio in
the target region (spaxel) are close to or differ from those
parameters in the calibrating points located close to the calibration
relation. We then determined abundance distributions within the
optical radii in the discs of 47 MaNGA galaxies. The optical radii of
the galaxies were estimated from the surface brightness profiles
constructed based on the MaNGA observations.
}


\titlerunning{Abundances based on spaxel spectra of the MaNGA survey}
\authorrunning{Pilyugin et al.}
\maketitle

\section{Introduction}

It is well established that the characteristic oxygen abundance of a
galaxy correlates with its macroscopic properties such as stellar
galaxy mass or luminosity \citep[][among many
others]{Lequeux1979,Zaritsky1994, Garnett2002, Grebel2003,
Tremonti2004, Erb2006, Cowie2008, Maiolino2008, Guseva2009, Thuan2010,
PilyuginThuan2011, Pilyugin2013, Andrews2013, Zahid2013, Maier2014,
Steidel2014, Izotov2015}.  On the other hand, whether or not a correlation
exists between the radial abundance distribution (gradient) and the
macroscopic parameters of a galaxy is still not clear.

Two-dimensional (2D) spectroscopy of a large number of nearby galaxies has
been carried out in the framework of surveys such as the Calar Alto
Legacy Integral Field Area (CALIFA) survey \citep{Sanchez2012a,
Husemann2013, GarciaBenito2015}  or the Mapping Nearby Galaxies at
Apache Point Observatory (MaNGA) survey \citep{Bundy2015}.   The 2D
spectroscopy provides the possibility to determine the abundances in
many regions (spaxels) across the visible surface of a galaxy and to
construct an abundance map. From such data, radial abundance gradients
in the discs of galaxies can be obtained.  The results of these
studies are somewhat contradictory. 

The radial oxygen abundance distributions across the CALIFA galaxies
were investigated by \citet{Sanchez2012b,Sanchez2014}, and
\citet{SanchezMenguiano2016}.  They found that all galaxies without
clear evidence of an interaction present a common gradient in the
oxygen abundance; the slope is independent of morphology, the
existence of a bar, the absolute magnitude, or mass, and the distribution
of the slopes is statistically compatible with a random Gaussian
distribution around a mean value.

\citet{Belfiore2017} measured the oxygen abundance gradients in a
sample of 550 nearby galaxies in the stellar mass range
$10^{9}$M$_{\odot}$ to $10^{11.5}$M$_{\odot}$ with spectroscopic data
from the SDSS-IV MaNGA survey. They found that the gradients in
galaxies of $\sim10^{9}$M$_{\odot}$ are roughly flat and that the
metallicity gradient steepens with stellar mass until
$\sim10^{10.5}$M$_{\odot}$, while it becomes flatter again for higher
masses.

Reliable values of the abundance gradients (and, consequently,
abundances) are essential to investigate the correlation between the
abundance gradient and other macroscopic properties of galaxies.
Determination of the abundances from the spaxel spectra encounters two
problems.  Firstly, there is no unique metallicity scale for the
H\,{\sc ii} regions since different methods used for the abundance
determinations produce the oxygen abundances that may differ
significantly.  Moreover, the majority of the widely used methods do
not work over the whole metallicity scale of the  H\,{\sc ii} regions.
Instead they are applicable for a limited range of metallicities only.
This can introduce an uncertainty even in the relative values of the
gradients determined with the same method in galaxies of different
masses.

Secondly, it is believed that not only the H\,{\sc ii} regions but
also other objects (e.g. diffuse ionized gas, (DIG)) can contribute to
the spaxel spectra of the MaNGA survey
\citep{Belfiore2015,Belfiore2017,Zhang2017,Sanders2017}.  The strength
of low ionization lines, [N\,{\sc ii}], [O\,{\sc ii}], and [S\,{\sc
ii}], may be increased relative to Balmer lines in (some) spaxel
spectra.  When this happens, the abundances derived from the spaxel
spectra through strong-line methods may suffer from large errors.

Before determining the abundance distributions across the discs of the
MaNGA galaxies, we examine the validity of the abundances obtained
through the strong-line methods from the spaxel spectra by comparing
the behaviour of the line strengths and the abundances estimated
through different calibrations for a sample of slit spectra of H\,{\sc
ii} regions in nearby galaxies, of fibre spectra from the SDSS, and of
spaxel spectra of the MaNGA survey.  We ascertain which
calibration produces the most reliable abundances from the spaxel
spectra. Then the abundance distributions within the optical radii in
the discs of the MaNGA galaxies are determined.

The paper is organised in the following way. The data are described in
Section 2. In Section 3 the behaviour of the line intensities in slit
spectra of H\,{\sc ii} regions in nearby galaxies, of fibre spectra
from the SDSS, and of spaxel spectra of the MaNGA survey are compared
and analysed. The abundances for those samples of spectra are
estimated through different calibrations and examined.  The abundance
distributions in the discs of 47 MaNGA galaxies are obtained in
Section 4. A discussion is given in Section 5 and  Section 6 contains a
brief summary.

Throughout the paper, we use the following standard notations 
for the line intensities: \\ 
$R_2  = I_{\rm [O\,II] \lambda 3727+ \lambda 3729} /I_{{\rm H}\beta }$,  \\
$N_2  = I_{\rm [N\,II] \lambda 6548+ \lambda 6584} /I_{{\rm H}\beta }$,  \\
$S_2  = I_{\rm [S\,II] \lambda 6717+ \lambda 6731} /I_{{\rm H}\beta }$,  \\
$R_3  = I_{{\rm [O\,III]} \lambda 4959+ \lambda 5007} /I_{{\rm H}\beta }$.  \\
Based on these definitions, the excitation parameter $P$ is expressed
as $P = R_{3}/R_{23} = R_{3}/(R_{2} + R_{3})$.   The notation
(O/H)$^{*}$ = 12 +log(O/H) is used in order to permit us to write
our equations in a compact way.

\section{Data for our sample of MaNGA galaxies}

The publicly available data obtained within the framework of the MaNGA
SDSS DR13 survey \citep{Albareti2016} form the basis of the current
study. We selected our initial sample of 150 star-forming galaxies by
visual examination of the images of the MaNGA galaxies with the goal
being to determine the chemical (abundance distribution), kinematical
(rotation curve), and photometrical (surface brightness profile)
properties of these galaxies.  Only galaxies for which the MaNGA
observations cover the entire or a large fraction of their visible
extent were selected.  

For each spaxel spectrum, the fluxes of the emission lines were
measured. The velocity of each region (spaxel) is estimated from the
measured wavelength of the H$\alpha$ line.  The surface brightness in
the SDSS $g$ and $r$ bands was obtained from broadband SDSS images
created from the data cube.  Our final sample of the MaNGA galaxies
comprises 47 galaxies for which the chemical, kinematical, and
photometrical properties are derived within the whole optical radius
$R_{25}$ or at least up to $\sim$0.8$R_{25}$.  The chemical properties
of those galaxies are considered in our current study; other
properties will be discussed in a forthcoming study.  The MaNGA
galaxies included in the final sample are listed in
Table~\ref{table:sample}.

\subsection{Emission line fluxes}

\begin{figure*}
\resizebox{1.00\hsize}{!}{\includegraphics[angle=000]{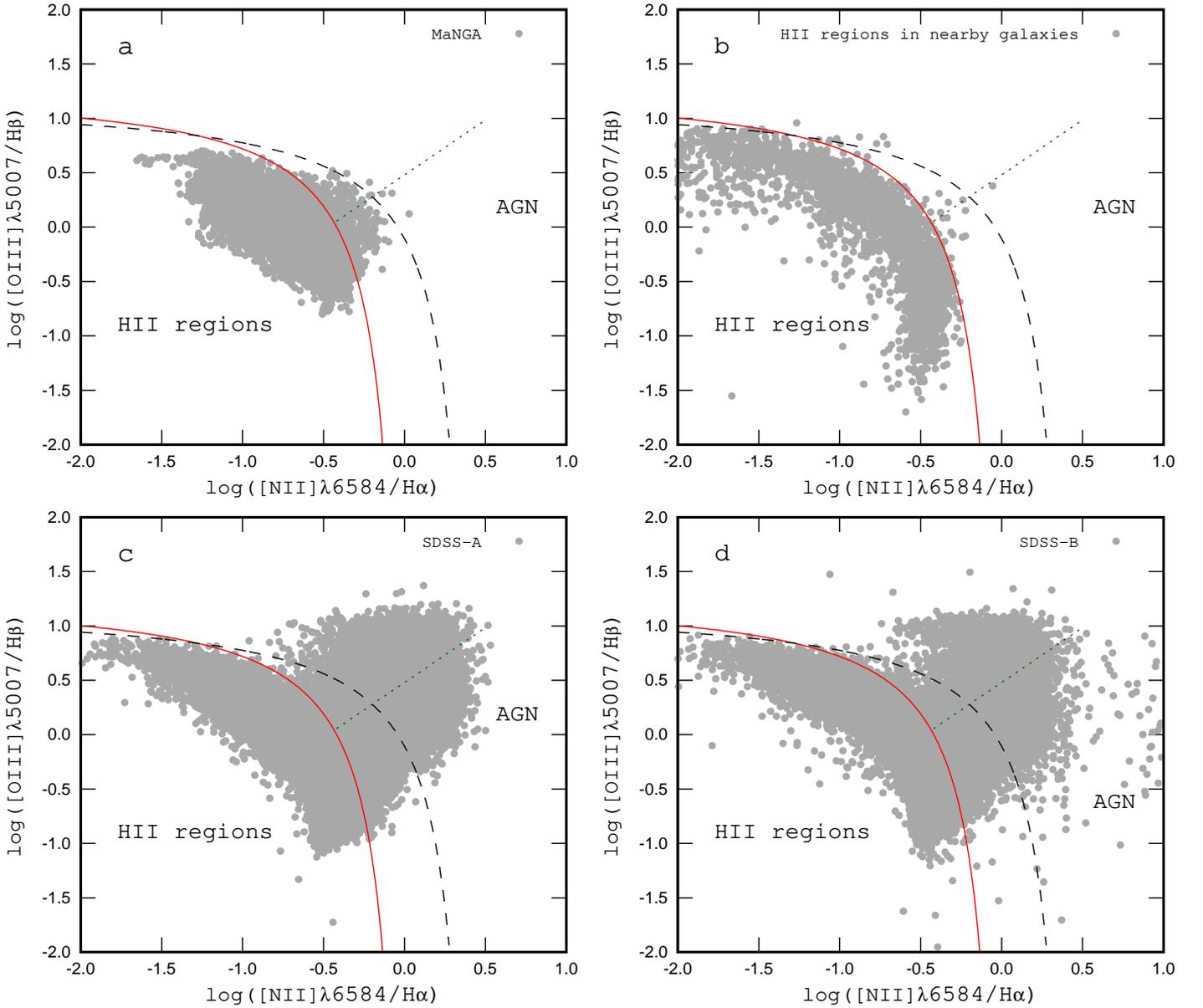}}
\caption{
Panel $a$ shows the BPT diagram for our sample of MaNGA galaxies.  The
filled circles are individual spaxels.  Solid and long-dashed curves
mark the demarcation line between AGNs and H\,{\sc ii} regions defined
by \citet{Kauffmann2003} and \citet{Kewley2001}, respectively.  The
short-dashed line is the dividing line between Seyfert galaxies and
LINERs defined by \citet{CidFernandes2010}.  Panel $b$ shows the BPT
diagram for  H\,{\sc ii} regions in nearby galaxies.  Panels $c$ and
$d$ are the BPT diagrams for fibre spectra of the SDSS galaxies with
line measurements from the catalogues SDSS-A and SDSS-B, respectively.
}
\label{figure:bpt}
\end{figure*}

\begin{figure}
\resizebox{1.00\hsize}{!}{\includegraphics[angle=000]{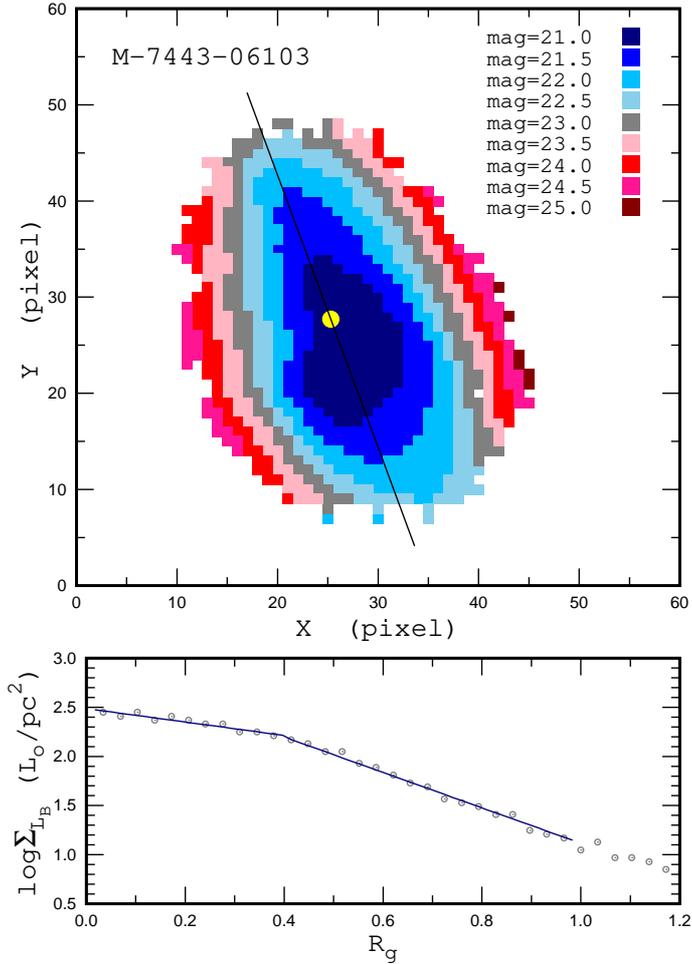}}
\caption{
The photometric properties of the MaNGA galaxy M-7443-06103.  The
upper panel shows the surface brightness distribution across the image
of the galaxy M-7443-06103 in sky coordinates (pixels). The value of
the surface brightness is colour-coded.  The circle shows the
kinematic centre of the galaxy.  The line indicates the position of
the major kinematic axis of the galaxy.  The lower panel shows the
observed surface-brightness profile (points) and the broken
exponential fit within the isophotal radius $R_{25}$ (line).
}
\label{figure:SBprofile-example}
\end{figure}

The spectrum of each spaxel from the MaNGA datacubes is reduced in the
manner described in \citet{Zinchenko2016}. For each spectrum, the
fluxes of the 
[O\,{\sc ii}]$\lambda$3727+$\lambda$3729, 
H$\beta$,  
[O\,{\sc iii}]$\lambda$4959, 
[O\,{\sc iii}]$\lambda$5007,
[N\,{\sc ii}]$\lambda$6548,
H$\alpha$,  
[N\,{\sc ii}]$\lambda$6584, 
[S\,{\sc ii}]$\lambda$6717, and 
[S\,{\sc ii}]$\lambda$6731 lines are measured. The velocity of each 
region (spaxel) is estimated from the measured wavelength of the 
H$\alpha$ line.  We selected spectra for which the signal-to-noise 
ratio is larger than three, S/N $>$ 3, for each of those lines.

The [O\,{\sc iii}]$\lambda$5007 and $\lambda$4959 lines originate from transitions from the 
same energy level, so their flux ratio is determined only by the transition probability ratio, 
which is  very close to 3 \citep{Storey2000}. Thus, the value of $R_3$ can be estimated as
$R_3  = I_{{\rm [O\,III]} \lambda 4959+ \lambda 5007} /I_{{\rm H}\beta }$ or as 
$R_3  = 1.33I_{{\rm [O\,III]} \lambda 5007} /I_{{\rm H}\beta }$. 
The stronger line [O\,{\sc iii}]$\lambda$5007 
is usually measured with higher precision than the weaker line [O\,{\sc iii}]$\lambda$4959. 
Therefore, we estimated the value of $R_3$ to be $R_3  = 1.33$~[O\,{\sc iii}]$\lambda$5007/H$\beta$
rather than using the sum of the line fluxes
[O\,{\sc iii}]$\lambda$5007 and $\lambda$4959.
Similarly, the [N\,{\sc ii}]$\lambda$6584 and $\lambda$6548 lines also originate from transitions from the 
same energy level and the transition probability ratio for those lines is again
close to 3 \citep{Storey2000}.
We estimated the value of $N_2$ to be $N_2 = 1.33$~[N\,{\sc
ii}]$\lambda$6584/H$\beta$ instead of the sum of the [N\,{\sc
ii}]$\lambda$6584 and $\lambda$6548 lines.

The lines [O\,{\sc ii}]$\lambda$3727+$\lambda$3729, 
H$\beta$,  
[O\,{\sc iii}]$\lambda$5007,
H$\alpha$,  
[N\,{\sc ii}]$\lambda$6584, 
[S\,{\sc ii}]$\lambda$6717, and
[S\,{\sc ii}]$\lambda$6731 
are used for the dereddening and abundance determinations.
Specifically, we corrected the emission-line fluxes for interstellar
reddening using the theoretical H$\alpha$/H$\beta$ ratio and the
\citet{Whitford1958} interstellar reddening law (adopting the
approximation suggested by \citet{Izotov1994}).

We used the standard diagnostic diagram of the [N\,{\sc
ii}]$\lambda$6584/H$\alpha$ versus the [O\,{\sc
iii}]$\lambda$5007/H$\beta$ line ratios suggested by
\citet{Baldwin1981}, which is known as the BPT classification diagram,
to separate H\,{\sc ii} region-like objects and active galactic nuclei (AGN)-like objects. We
adopted the demarcation line of \citet{Kauffmann2003} between H\,{\sc
ii} regions and AGNs.  The BPT diagram for our sample of the MaNGA
galaxies is presented in Panel $a$ of Fig.~\ref{figure:bpt}. The
spectra of 48591 spaxels in 47 MaNGA galaxies are used to determine
the abundance distributions in the discs of those galaxies.

\subsection{Galaxy size}

\subsubsection{Preliminary remarks}

The most basic structural property of a galaxy is its size. The
definition and the estimation of the galaxy size is not a
 trivial task. Different values (isophotal or optical
radius, effective radius, Petrosian radius) have been suggested and
used to specify the galaxy size.

The isophotal or optical radius is measured at or reduced
to a fixed surface brightness level and corrected to a ``face-on''
position of a galaxy. Two values of the surface brightness level are
widely used, that is, the standard isophotal radius $R_{25}$, which
refers to a level of 25.0 mag/arcsec$^{2}$ in the B-band listed in
the Third Reference Catalog of Bright Galaxies  \citep{RC3}, and the
Holmberg radius, which refers to the Holmberg isophote at a level of
26.5 mag/arcsec$^{2}$  given in the Updated Nearby Galaxy Catalog
\citep{Karachentsev2013}.  Other values of the surface brightness
level (and other photometric bands) are also used. 

The  effective radius is defined by the total magnitude of a
galaxy.  The total (asymptotic) magnitude of a galaxy is estimated by
extrapolating the growth curve (the integrated magnitude as a function
of radius) to infinity. The effective radius of a galaxy $R_{eff}$ is
defined as the radius containing half of the total luminosity. The
effective radius depends on the level of central concentration, in the
sense that highly concentrated galaxies tend to have small values of
$R_{eff}$ compared to the overall extent of the galaxy
\citep{MunozMateos2015}. 

The Petrosian radius is estimated through the inverse Petrosian
index, which is the ratio of the surface brightness at a given radius
to the average surface brightness inside this radius
\citep{Petrosian1976}.  The inverse Petrosian index is estimated at
each successively increasing radius. When the ratio falls below a
chosen small value (usually adopted to be 0.2), that radius is
multiplied by a factor (usually adopted to be 1.5) to obtain the
Petrosian radius.

There is no one-to-one correspondence between those three values of
the galaxy size. Here we use the standard isophotal radius
$R_{25}$ to specify the size of the galaxy. The fractional
galactocentric distances of the spaxels $R_{g}$ are normalized to
the optical radius $R_{25}$. 

It is interesting to note that the investigations of the extended
discs of galaxies can provide a physically justified definition of the
edge of a galaxy. It was found
\citep{Bresolin2009,Goddard2011,Bresolin2012,Patterson2012,Kudritzki2014,SanchezMenguiano2016}
that the radial abundance gradient flattens to an approximately
constant value outside of the isophotal radius. \citet{Bresolin2012}
and \citet{Kudritzki2014}  concluded that it is unlikely that the
in-situ star formation outside of the isophotal radius could have
produced and ejected enough metals to enrich the interstellar medium
to the presently observed values.  If this is a universal property of
spiral galaxies then one can define the edge of a galaxy as the
largest radius where the observed abundance can be produced by the
stars located there.

\subsubsection{Determinations of optical radii}

The surface brightness in the SDSS $g$ and $r$ bands for each spaxel
was obtained from broadband SDSS images created from the data cube. 
The measurements were corrected for Galactic
foreground extinction using the recalibrated $A_V$ values of
\citet{Schlafly2011} taken from the NASA/IPAC
Extragalactic Database ({\sc ned})\footnote{The NASA/IPAC
Extragalactic Database ({\sc ned}) is operated by the Jet Populsion
Laboratory, California Institute of Technology, under contract with
the National Aeronautics and Space Administration.  {\tt
http://ned.ipac.caltech.edu/} }.
Then the measurements in the {\it SDSS} filters $g$ and $r$
were converted to $B$-band magnitudes using the conversion relations
of \citet{Blanton2007},
\begin{equation} 
B_{AB} = g + 0.2354 + 0.3915\;[(g-r)-0.6102] ,
\label{equation:bgr} 
\end{equation} 
where the $B_{AB}$, $g$, and $r$ magnitudes in
Eq.~(\ref{equation:bgr}) are in the $AB$ photometric system.  The $AB$
magnitudes were reduced to the Vega photometric system 
\begin{equation} 
B_{Vega} = B_{AB} + 0.09  
\label{equation:VegaAB} 
,\end{equation} 
which was done using the relation of \citet{Blanton2007}.
The surface-brightness profile was determined from the surface brightness
map with the galaxy inclination and the position angle of the major
axis obtained through the analysis of the observed velocity field.
Those values will be discussed in a forthcoming paper.  The upper
panel of Fig.~\ref{figure:SBprofile-example} shows the
surface-brightness distribution across the image of the galaxy
M-7443-06103 in sky coordinates (pixels).  The position of the
kinematic centre of the galaxy is marked by the circle and the major
kinematic axis of the galaxy is indicated by the line in the upper
panel of Fig.~\ref{figure:SBprofile-example}.  The lower panel shows
the observed surface-brightness profile (points) for the galaxy
M-7443-06103. 

We use the surface-brightness profile in solar units for the
bulge-disc decomposition. The magnitude of the Sun in the $B$ band of
the Vega photometric system,  $B_{\sun}$ = 5.45, was taken from
\citet{Blanton2007}.  The distances were adopted from the {\sc ned}.
The {\sc ned} distances use flow
corrections for Virgo, the Great Attractor, and Shapley Supercluster
infall.  The stellar surface-brightness distribution within a galaxy
was fitted by a broken exponential profile for the disc and by a
general S\'{e}rsic profile for the bulge. 

The total surface-brightness distribution was fitted with the expression 
\begin{eqnarray}
       \begin{array}{lll}
\Sigma_L(r) & = & (\Sigma_L)_{e}\exp \{-b_{n}[(r/r_{e})^{1/n} - 1]\} \\
            & + & (\Sigma_{L})_{0,inner}\exp(-r/h_{inner}) \;\;\;\; if \;\;\; r < R^{*} ,                        \\
            & = & (\Sigma_L)_{e}\exp \{-b_{n}[(r/r_{e})^{1/n} - 1]\} \\
            & + & (\Sigma_{L})_{0,outer}\exp(-r/h_{outer}) \;\;\;\; if \;\;\; r > R^{*}  .                        \\
     \end{array}
\label{equation:decomp2}
\end{eqnarray}
Here $R^{*}$ is the break radius, that is, the radius at which the
exponent changes.  The eight parameters $(\Sigma_{L})_{e}$, $r_{e}$,
$n$, $(\Sigma_{L})_{0,inner}$, $h_{inner}$, $(\Sigma_{L})_{0,outer}$,
$h_{outer}$, and $R^{*}$ in the broken exponential disc were
determined through the best fit to the observed surface-brightness
profile, that is, we require that the deviation $\sigma_{BED}$ , given by
\begin{equation}
\sigma = \sqrt{ [\sum\limits_{j=1}^n (L(r_{j})^{cal}/L(r_{j})^{obs} - 1)^2]/n}  ,
\label{equation:sigma}
\end{equation}
be minimized. Here $L(r_{j})^{cal}$ is the surface brightness at the
radius $r_{j}$ computed through Eq.~(\ref{equation:decomp2}) and
$L(r_{j})^{obs}$ is the measured surface brightness at that radius.
The broken exponential fit to the surface-brightness profile of the
galaxy M-7443-06103 is shown by the line in the lower panel of
Fig.~\ref{figure:SBprofile-example}. 

The isophotal radius of the galaxy, the central surface brightness of
the disc, and the galaxy luminosity estimated using 
Eq.~(\ref{equation:decomp2}) are reported in Table~\ref{table:sample}.

\begin{table*}
\caption[]{\label{table:sample}
The characteristics of MaNGA galaxies of our sample.
The first column gives the galaxy's identification number in the MaNGA survey.
The right ascension (R.A.) and declination (Dec.) (J2000.0) of each galaxy are given in
columns 2 and 3.
The adopted distance is reported in column 4.  
The isophotal radius in arcmin and in kpc are listed in columns 5 and
6, respectively. 
The luminosity and the central surface brightness of the disc (reduced to the face-on) in the $B$ band 
estimated here are given in columns 7 and 8.
The central oxygen abundance and the radial abundance gradient estimated using the $R$ calibration
are reported in columns 9 and 10.
Column 11 gives the value of the mean scatter around the O/H -- $R_{g}$ relation.
}
\begin{center}
\begin{tabular}{ccccccccccc} \hline \hline
  Name                       &
R.A.                         &
Dec.                         &           
d                            &
R$_{25}$                      &
R$_{25}$                      &
logL$_{B}$                    & 
log$\Sigma_{0}$               &
12+log(O/H)$_{0}$             &
gradient O/H                 &
$\sigma$ O/H                 \\
                             &
$\degr$                      &
$\degr$                      &
Mpc                          &
arcmin                       &
kpc                          &
L$_{\odot}$                   &  
L$_{\odot}$/pc$^{2}$           &
                             &
dex/R$_{25}$                  &
dex                          \\  \hline
 7443 06103 &  230.233668 &   42.775541 &   82.8 &  0.24 &   5.82 &   9.61 &   2.20 &  8.41 &  -0.081 &  0.0547 \\ 
 7957 06104 &  258.220415 &   35.093157 &  111.2 &  0.27 &   8.63 &  10.00 &   2.31 &  8.47 &  -0.173 &  0.0535 \\ 
 7977 06103 &  331.802634 &   13.266052 &  143.5 &  0.16 &   6.61 &   9.34 &   1.69 &  8.53 &  -0.325 &  0.0554 \\ 
 8131 12701 &  111.109940 &   38.945215 &   72.5 &  0.36 &   7.56 &   9.56 &   2.04 &  8.46 &  -0.379 &  0.0680 \\ 
 8134 12701 &  112.698680 &   46.010115 &   88.9 &  0.32 &   8.19 &   9.51 &   1.84 &  8.54 &  -0.309 &  0.0669 \\ 
 8134 12703 &  114.473654 &   47.957220 &   88.1 &  0.30 &   7.69 &   9.42 &   1.71 &  8.52 &  -0.231 &  0.0708 \\ 
 8134 12705 &  114.673017 &   45.943654 &   76.4 &  0.33 &   7.41 &   9.41 &   1.77 &  8.59 &  -0.235 &  0.0552 \\ 
 8135 03704 &  114.897365 &   37.751505 &  126.7 &  0.28 &  10.13 &  10.06 &   2.58 &  8.48 &  -0.228 &  0.0564 \\ 
 8138 06104 &  117.291419 &   46.461883 &  158.4 &  0.15 &   6.91 &   9.49 &   1.77 &  8.47 &  -0.175 &  0.0579 \\ 
 8140 09101 &  116.507218 &   40.875058 &  157.9 &  0.17 &   8.04 &   9.66 &   2.06 &  8.43 &  -0.245 &  0.0696 \\ 
 8143 06102 &  118.946946 &   41.280373 &  170.1 &  0.21 &  10.31 &   9.79 &   1.93 &  8.46 &  -0.275 &  0.0711 \\ 
 8243 09101 &  128.178383 &   52.416775 &  179.8 &  0.27 &  13.95 &  10.22 &   2.34 &  8.44 &  -0.208 &  0.0542 \\ 
 8247 12701 &  138.275133 &   43.013463 &   62.1 &  0.31 &   5.57 &   9.11 &   1.63 &  8.24 &  -0.190 &  0.0664 \\ 
 8250 12702 &  140.359627 &   43.102154 &  188.9 &  0.22 &  11.91 &   9.91 &   1.96 &  8.62 &  -0.327 &  0.0494 \\ 
 8250 12703 &  139.647743 &   44.596737 &   62.9 &  0.28 &   5.03 &   8.96 &   1.46 &  8.25 &  -0.216 &  0.0798 \\ 
 8252 12703 &  145.456567 &   48.648991 &  189.8 &  0.20 &  11.04 &   9.87 &   2.10 &  8.64 &  -0.247 &  0.0478 \\ 
 8253 03703 &  157.343576 &   43.170589 &  116.7 &  0.20 &   6.79 &   9.88 &   2.60 &  8.59 &  -0.019 &  0.0181 \\ 
 8253 09102 &  158.207632 &   43.173946 &  124.4 &  0.38 &  13.57 &  10.11 &   2.17 &  8.52 &  -0.309 &  0.0456 \\ 
 8257 03704 &  165.553613 &   45.303871 &   88.8 &  0.24 &   6.24 &   9.52 &   2.04 &  8.27 &  -0.179 &  0.0647 \\ 
 8258 03701 &  165.570030 &   44.335183 &  106.4 &  0.19 &   5.93 &   9.71 &   2.86 &  8.63 &  -0.310 &  0.0458 \\ 
 8258 06101 &  168.015359 &   42.669794 &   93.3 &  0.25 &   6.78 &   9.37 &   1.80 &  8.27 &  -0.143 &  0.0618 \\ 
 8259 12701 &  179.962635 &   43.740689 &   87.1 &  0.32 &   8.02 &   9.48 &   1.71 &  8.32 &  -0.194 &  0.0657 \\ 
 8259 12702 &  178.506318 &   44.642256 &  104.5 &  0.38 &  11.40 &   9.85 &   1.87 &  8.62 &  -0.256 &  0.0435 \\ 
 8317 12705 &  193.855952 &   43.938063 &  165.3 &  0.21 &  10.02 &   9.61 &   1.64 &  8.50 &  -0.216 &  0.0739 \\ 
 8318 12705 &  197.411496 &   45.269246 &  125.8 &  0.33 &  11.89 &  10.15 &   2.22 &  8.65 &  -0.366 &  0.0410 \\ 
 8320 12701 &  205.548609 &   22.269154 &  180.3 &  0.19 &  10.05 &   9.70 &   2.00 &  8.52 &  -0.246 &  0.0593 \\ 
 8320 12703 &  206.630975 &   23.122136 &  132.7 &  0.26 &   9.97 &   9.66 &   1.72 &  8.53 &  -0.272 &  0.0473 \\ 
 8325 12701 &  210.176336 &   45.833422 &  182.9 &  0.26 &  13.74 &   9.88 &   1.77 &  8.57 &  -0.256 &  0.0631 \\ 
 8325 12705 &  212.101063 &   47.273720 &  168.4 &  0.21 &  10.21 &   9.76 &   1.88 &  8.61 &  -0.306 &  0.0507 \\ 
 8329 12704 &  213.783652 &   45.594849 &   72.9 &  0.42 &   9.01 &   9.86 &   2.23 &  8.47 &  -0.304 &  0.0564 \\ 
 8335 06102 &  216.904792 &   40.407305 &   86.8 &  0.28 &   7.15 &   9.59 &   2.40 &  8.42 &  -0.218 &  0.0606 \\ 
 8341 09102 &  191.012464 &   45.129265 &  107.4 &  0.24 &   7.55 &   9.68 &   1.99 &  8.58 &  -0.146 &  0.0394 \\ 
 8439 06101 &  143.184618 &   48.796348 &  111.9 &  0.25 &   8.14 &   9.76 &   2.17 &  8.57 &  -0.265 &  0.0511 \\ 
 8439 09101 &  142.713904 &   50.318861 &  159.0 &  0.13 &   6.17 &   9.37 &   1.79 &  8.49 &  -0.220 &  0.0537 \\ 
 8439 12701 &  143.010196 &   48.551093 &   71.4 &  0.37 &   7.62 &   9.26 &   1.61 &  8.43 &  -0.257 &  0.0740 \\ 
 8440 12703 &  136.748725 &   41.440882 &  110.8 &  0.27 &   8.59 &   9.65 &   1.90 &  8.51 &  -0.311 &  0.0546 \\ 
 8447 12703 &  207.483361 &   40.895076 &  150.8 &  0.12 &   5.12 &   8.89 &   1.26 &  8.47 &  -0.219 &  0.0738 \\ 
 8451 12703 &  164.028900 &   43.156568 &  160.4 &  0.25 &  11.66 &   9.86 &   1.99 &  8.41 &  -0.187 &  0.0742 \\ 
 8453 12702 &  151.547771 &   47.295038 &  160.7 &  0.21 &   9.74 &   9.56 &   1.51 &  8.53 &  -0.216 &  0.0665 \\ 
 8459 06102 &  147.990674 &   43.414043 &   70.1 &  0.23 &   4.59 &   9.20 &   2.13 &  8.33 &  -0.171 &  0.0544 \\ 
 8459 12704 &  147.604836 &   44.040637 &   70.0 &  0.23 &   4.75 &   8.92 &   1.59 &  8.48 &  -0.232 &  0.0651 \\ 
 8459 12705 &  148.117076 &   42.819141 &   71.9 &  0.38 &   7.84 &   9.48 &   1.71 &  8.52 &  -0.294 &  0.0595 \\ 
 8466 09101 &  170.214847 &   44.597411 &  106.8 &  0.27 &   8.28 &   9.57 &   1.74 &  8.56 &  -0.289 &  0.0591 \\ 
 8466 09102 &  170.583304 &   46.701332 &   77.8 &  0.28 &   6.41 &   9.53 &   2.07 &  8.28 &  -0.210 &  0.0565 \\ 
 8485 12701 &  233.319217 &   48.119650 &  102.7 &  0.28 &   8.22 &   9.52 &   1.82 &  8.34 &  -0.189 &  0.0602 \\ 
 8548 06103 &  243.472257 &   48.296700 &   89.9 &  0.33 &   8.72 &   9.82 &   2.26 &  8.45 &  -0.255 &  0.0396 \\ 
 8549 06103 &  240.418740 &   46.085291 &  177.2 &  0.23 &  11.60 &   9.82 &   1.91 &  8.43 &  -0.200 &  0.0560 \\ 
\hline
\end{tabular}\\
\end{center}
\begin{flushleft}
\end{flushleft}
\end{table*}

It should be noted that the bulge contribution is small in each of the
galaxies considered, and the stellar surface-brightness distribution
within a galaxy is well-fitted by a broken exponential profile for the
disc, that is, our sample of MaNGA galaxies can be considered as a sample
of bulgeless galaxies.

\subsection{Abundances}

\subsubsection{Calibrations used}

It is believed that the radiation of the diffuse ionized gas can
contribute significantly to (some) spaxel spectra of the MaNGA survey,
and the strength of the low-ionization lines [N\,{\sc ii}], [O\,{\sc
ii}], and [S\,{\sc ii}] can be increased relative to the Balmer lines.
As a result, the abundances derived from the spaxel spectra through
strong-line methods may have large errors
\citep{Belfiore2015,Belfiore2017,Zhang2017,Sanders2017}.
We consider four calibrations based on different low-ionization emission
lines to estimate the oxygen abundances from the spaxel spectra.  On
the one hand, the use of the calibrations based on different emission
lines allows us to draw conclusions about the validity of the abundances
and to choose the calibration that produces the most reliable
abundances.  On the other hand, this provides the possibility to
establish which of the low-ionization emission lines are significantly
distorted by the contribution of the radiation of the diffuse ionized
gas to the spaxel spectra.

We use the simple three-dimensional (3D) calibration relations for
abundance determinations from a set of strong emission lines suggested
in \citet{Pilyugin2016}. The oxygen abundances (O/H)$_{R}$ are
determined using the oxygen R$_2$, R$_3$ lines and the nitrogen N$_2$
line ($R$ calibration).  The oxygen R$_3$ line, the nitrogen N$_2$ line,
and the sulphur S$_2$ line (instead of the oxygen R$_2$ line) are used in the
determinations of the (O/H)$_{S}$ abundances through the $S$
calibration introduced in \citet{Pilyugin2016}. 

The 2D $P$ calibration is based on the oxygen R$_2$ and
R$_3$ lines only \citep{Pilyugin2000,Pilyugin2001,Pilyugin2005}.  It
is important that the calibration relations were consistent with each
other.  We obtained the $P$ calibration relation for high-metallicity
objects, 12 +log(O/H) $\ga 8.35$, (the so-called upper-branch
calibration) using the (O/H)$_{R}$ abundances for the calibrating data
points (the compilation of spectra of H\,{\sc ii} regions in nearby
galaxies) from \citet{Pilyugin2012,Pilyugin2014}, and
\citet{Pilyugin2016}).  The obtained upper-branch $P$ calibration
relation is 
\begin{eqnarray}
       \begin{array}{lll}
 {\rm (O/H)}^{*}_{P}  & = &  7.852 + 0.332 \, P  \\  
                     & + & (-0.560 - 0.176 \, P)  \times (\log R_{3} -1)  \\ 
                     & + & (-0.120 + 0.109 \, P)  \times (\log R_{3} -1)^{2}  \\ 
     \end{array}
\label{equation:ohp}
,\end{eqnarray}
where $P$ is the excitation parameter. The notation (O/H)$^{*}_{P}$ =
12 +log(O/H)$_{P}$ is used for the sake of brevity. 

The nitrogen N$_2$ line is the basis of the one-dimensional (1D) N2
calibration of \citet{Pettini2004}.  We obtained the calibration
relation for objects with logN$_2 > -0.6$ using again the (O/H)$_{R}$
abundances for the calibrating data points 
\begin{equation}
12 + \log{\rm (O/H)}_{N} = 8.485 + 0.616 \, \log N_{2}  .
\label{equation:ohn}
\end{equation}
It should be noted that the nitrogen N$_2$ line is used in this
relation but not the N2 abundance indicator of \citet{Pettini2004}
(see below).

Thus, we use four calibrations based on emission lines of different
low-ionization ions.  Only the nitrogen line N$_{2}$ is used in the
determinations of the (O/H)$_N$ abundances, only the oxygen line
R$_{2}$ is used in the determinations of the (O/H)$_P$ abundances, the
oxygen R$_{2}$ and nitrogen N$_{2}$ lines are used in the
determinations of the (O/H)$_R$ abundances, and the nitrogen N$_{2}$ and
sulphur S$_{2}$ lines are used in the determinations of the (O/H)$_S$
abundances.

\subsubsection{Abundance gradients}

The radial oxygen abundance distribution in a galaxy is traditionally
described by the linear relation 
\begin{equation}
12 +  \log {\rm (O/H)}(R) = 12 +  \log {\rm (O/H)}_{0} + grad \, \times \, R
\label{equation:grad}
,\end{equation}
where 12 + log(O/H)($R$) is the oxygen abundance at the galactocentric
distance $R$ in kpc, 12 + log(O/H)$_{0}$ is the extrapolated central
oxygen abundance, and $grad$ is the slope of the oxygen-abundance
gradient expressed in terms of dex/kpc.  The linear relation is
a rather good approximation for the abundance distributions in the
discs of spiral galaxies \citep[e.g.][and discussion
there]{Pilyugin2017}.  This is a standard physical scale where the
value of the gradient specifies the change of the abundance in the
disc as a function of the change of the radius in kiloparsecs.  

There is also another widely used definition of the gradient where the
change of the abundance refers not to the kpc scale but to some fixed
scale of galactocentric distance.  The gradient is often defined as
the change of the abundance in the disc over its isophotal radius
$R_{25}$ and is expressed in terms of dex/$R_{25}$.  In this
case, the galactocentric distance is normalized to the disc's
isophotal radius, $R_{g}$ = $R/R_{25}$.  With this definition, the
value of the gradient means the difference between the abundance at
the optical radius of a galaxy and the central abundance.  

Moreover, in a number of investigations, the gradient is defined as
the change of the abundance in the disc as a function of  its
effective radius and is expressed in terms of dex/$R_{eff}$.  In
this case, the galactocentric distance is normalized to the disc's
effective radius, $R^{*}$ = $R/R_{eff}$.  

It is evident that 
\begin{equation}
grad_{R_{\rm scale}}  = R_{\rm scale}  \, \times \,grad_{\rm kpc} 
\label{equation:gg}
,\end{equation}
where $R_{\rm scale}$ = $R_{25}$ or $R_{\rm scale}$ = $R_{eff}$. 
$R_{\rm scale} = 1$ for the canonical definition of the gradient.

We will use below the definition of the gradient where the change of
the abundance refers to the isophotal radius $R_{25}$ of a galaxy and
is expressed in terms of dex/$R_{25}$.  The radial oxygen
abundance distribution in each galaxy of our sample is fitted by a
linear relation.  If there were points that showed large deviations from
the O/H -- $R_{g}$ relation ($d_{OH}$ $>$ 0.2 dex) then those points
were not used in deriving the final relations and were excluded from
further analysis.  It should be noted that points with such large
deviations are present only in a few galaxies and those points are not
very numerous.  The mean deviation from the final relations (the mean
value of the residuals of the relations) is given by 
\begin{equation}
\sigma_{OH} = \left(\frac{1}{n}\sum_{j=1}^{n} (\log ({\rm O/H})^{OBS}_{j} - \log ({\rm O/H})^{CAL}_{j})^{2} \right)^{1/2}  
\label{equation:delta}
,\end{equation}
and is usually around 0.04 dex to 0.07 dex; see
Table~\ref{table:sample}. Thus, the deviations of the rejected points
from the O/H -- $R_{g}$ relation exceed 3--5 $\sigma_{\rm OH}$.

\section{Abundances derived from H\,{\sc ii} region slit spectra, from 
SDSS fibre spectra, and from MaNGA spaxel spectra}

\begin{figure}
\resizebox{1.00\hsize}{!}{\includegraphics[angle=000]{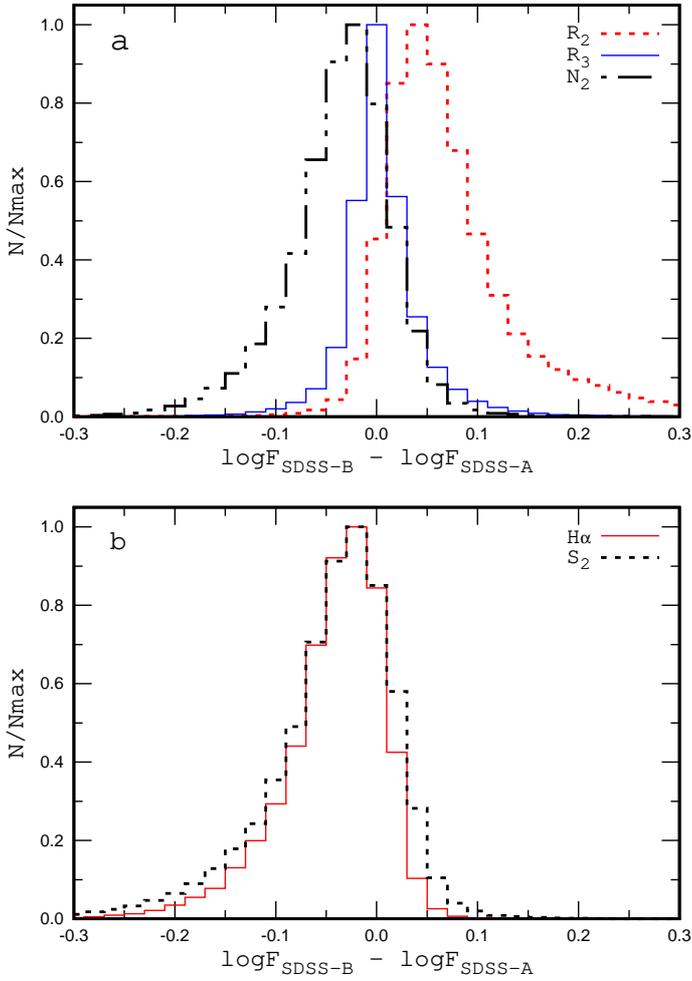}}
\caption{
The normalized histograms of the differences between the measured (non-dereddened)
emission-line intensities in the SDSS-A and SDSS-B catalogues of the SDSS fibre
spectra measurements. 
}
\label{figure:gist-dline}
\end{figure}

\begin{figure}
\resizebox{1.00\hsize}{!}{\includegraphics[angle=000]{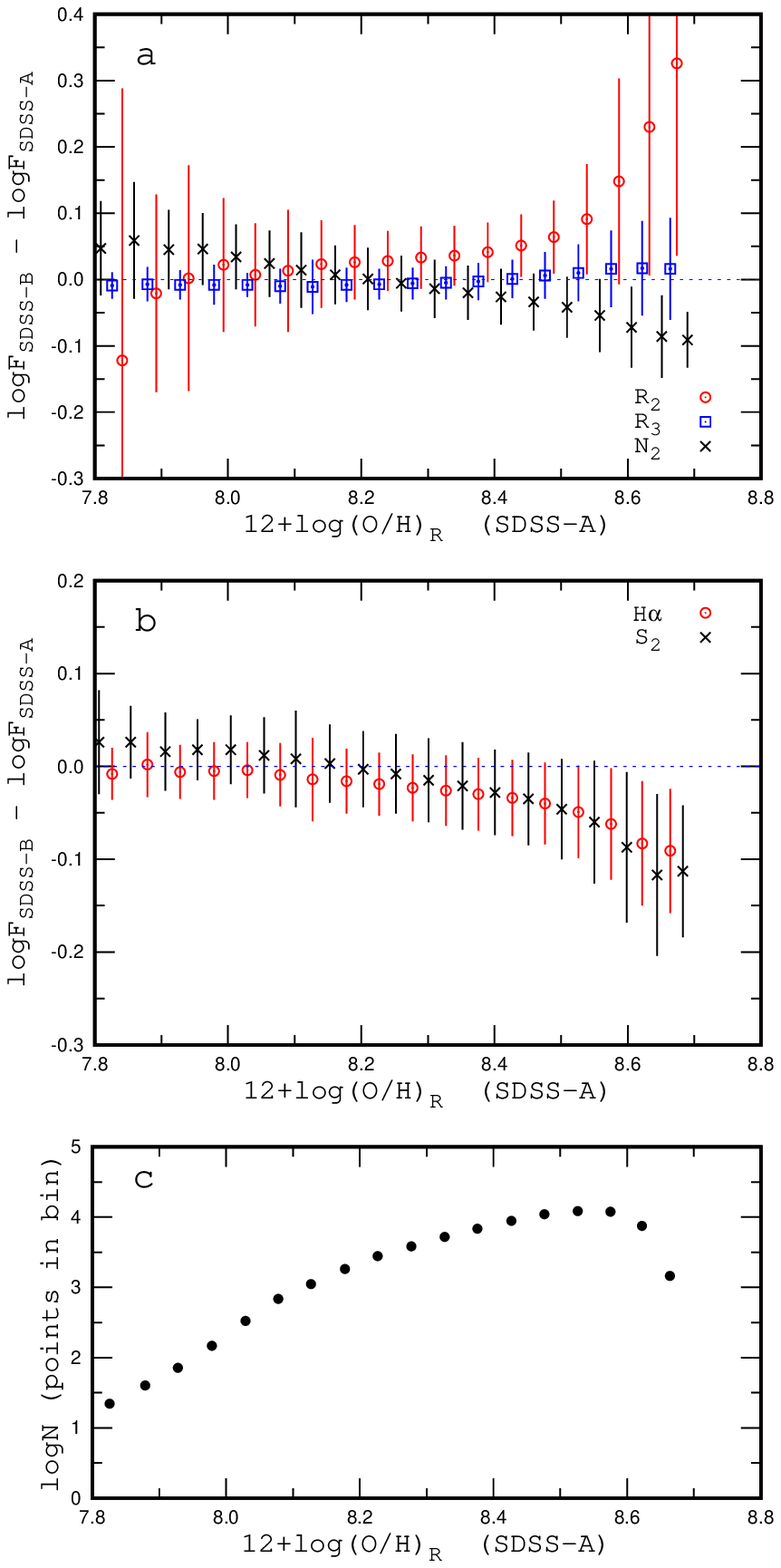}}
\caption{
Panels $a$ and $b$ show the differences between the measured
(non-dereddened) emission-line intensities in the SDSS-A and SDSS-B
catalogues. The symbols represent the mean values of the differences
for objects in bins of 0.05 dex in (O/H)$_{R,SDSS-A}$. The bars denote the
mean values of the scatter of the differences for objects in bins.
Panel $d$ shows the number of points (objects) in the bin. 
}
\label{figure:dline-ohr}
\end{figure}

\begin{figure}
\resizebox{1.00\hsize}{!}{\includegraphics[angle=000]{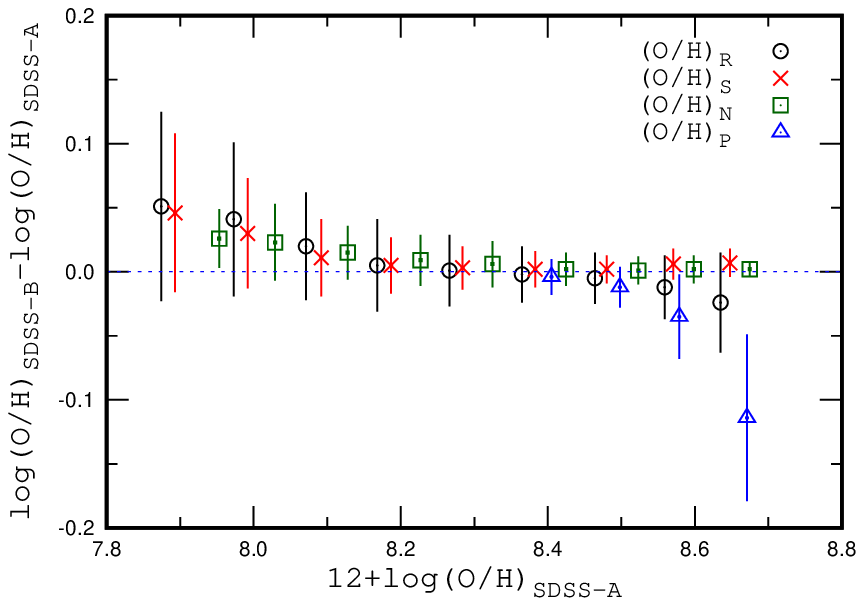}}
\caption{
The comparison of the oxygen abundances of the SDSS objects determined
through the different calibrations using the emission-line flux
measurements from the SDSS-A and SDSS-B catalogues. The symbols
represent the mean values of the abundance differences for objects
in bins of 0.1 dex in (O/H)$_{SDSS-A}$. The bars denote the mean values
of the scatter of the abundance differences for objects in the bins. 
}
\label{figure:ohmpa-doh}
\end{figure}

It is currently believed that the radiation of diffuse ionized gas can
make a significant contribution to (some) observed fibre spectra of
the SDSS survey and (some) spaxel spectra of the MaNGA survey and that
the strength of the low-ionization lines, [N\,{\sc ii}], [O\,{\sc ii}],
and [S\,{\sc ii}], can be increased relative to the Balmer lines
\citep[e.g.][]{Tremonti2004,Belfiore2015,Belfiore2017,Zhang2017,Sanders2017}.

Therefore, the abundances derived using strong-line methods from
spectra contaminated by the radiation of diffuse ionized gas can show large errors.
\citet{Zhang2017} found that the use of some
calibrations to derive metallicities introduce a bias in the derived
gradients as large as the gradient itself.  We have examined the
abundances estimated from slit spectra of a large sample of H\,{\sc
ii} regions in nearby galaxies, from fibre spectra of SDSS galaxies,
and from spaxel spectra of MaNGA galaxies. The abundances were
determined through different calibrations (based on different emission
lines).  If the radiation of the diffuse ionized gas produces a
distortion of the different low-ionization emission lines in the MaNGA
spaxel spectra \citep[e.g.][]{Sanders2017} then the abundances
determined using those lines as the abundance indicator will be
distorted to different extents. The comparison of abundances
determined through different calibrations allows us to estimate the
influence of the diffuse ionized gas on these determinations and to choose the calibration(s) producing
the most reliable abundances from the spaxel spectra of the MaNGA
galaxies.

\subsection{Samples considered} 

In our previous studies we compiled a large number of slit spectra of H\,{\sc ii} regions
in nearby spiral and irregular galaxies 
\citep{Pilyugin2012,Pilyugin2014,Pilyugin2016}.  The BPT diagram for
our sample of H\,{\sc ii} regions is presented in Panel $b$ of
Fig.~\ref{figure:bpt}. Some spectra located to the right of (or above)
the demarcation line of \citet{Kauffmann2003} were rejected. The final
sample of the  H\,{\sc ii} regions includes 3440 spectra with measured
R$_{2}$, R$_{3}$, N$_{2}$, and S$_{2}$ lines.

Two samples of emission-line fluxes in the SDSS fibre spectra are
considered.  Firstly, the emission-line flux measurements in the SDSS
fibre spectra were taken from the MPA/JHU catalogue
({\it galSpecLine}, SDSS-III DR12)
of automatic line flux measurements of the SDSS spectra (see
\citet{Brinchmann2004,Tremonti2004} and other publications of those
authors).  The stellar population models from \citet{Bruzual2003} are
used for the continuum fitting in this catalogue.  We select from the
catalogue those objects for which the S/N is larger than 3 for each of
the H$_{\beta}$,  H$_{\alpha}$, R$_{2}$, R$_{3}$, N$_{2}$, and S$_{2}$
lines in the spectra, and that are located left of (below) the
demarcation line of \citet{Kauffmann2003} in the BPT diagram in Panel
$c$ of Fig.~\ref{figure:bpt}. These criteria yield a sample of 193.037
SDSS objects, which we call the SDSS-A sample.

Secondly, we also used emission-line flux measurements for the SDSS
fibre spectra from the catalogue ({\it emissionLinesPort}, SDSS-III DR12).
The stellar population models
from \citet{Maraston2011} and \citet{Thomas2011} are used for the
continuum fitting in this catalogue.  The same criteria as for the
SDSS-A sample are used to select the objects from this catalogue. This
sample involves 87.381 SDSS objects and is called SDSS-B sample below.
The BPT diagram for SDSS-B sample is presented in Panel $d$ of
Fig.~\ref{figure:bpt}.

The MaNGA sample includes our measurements of the 48.591 spaxel
spectra in 47 MaNGA galaxies (see above).

\subsection{Uncertainty in the SDSS-fibre-based abundances due to the 
uncertainty in the data reduction and emission-line measurements} 

The samples SDSS-A and SDSS-B have 75.967 objects in common. This
provides the possibility to assess the uncertainties of the data
reduction and emission-line flux measurements in the SDSS fibre
spectra and corresponding uncertainties in the abundances determined
through the different calibrations.

Figure~\ref{figure:gist-dline} shows the normalized histograms of the
differences of the measured (non-dereddened) emission-line intensities
in the SDSS-A and SDSS-B catalogues of the SDSS fibre spectra.
Inspection of Fig.~\ref{figure:gist-dline} reveals
systematic differences between the emission-line intensities in the
SDSS-A and SDSS-B catalogues. The difference is dependent on the
wavelength of the emission line, in the sense that the measured
intensities of the ultraviolet line R$_{2}$ are stronger in the SDSS-B
catalogue than in the SDSS-A catalogue while the red lines, N$_{2}$,
H${\alpha}$, and N$_{2}$, are stronger in the SDSS-A catalogue than in the
SDSS-B catalogue.

Panels $a$ and $b$ of Fig.~\ref{figure:dline-ohr} show the differences
of the measured emission-line intensities between the SDSS-A and SDSS-B
catalogues of the SDSS fibre spectra as a function of oxygen abundance
(O/H)$_{R}$ estimated from the line measurements from the SDSS-A
catalogue.  The symbols represent the mean values of the differences
 for objects 
in bins of 0.05 dex in (O/H)$_{R}$. The bars denote the mean values of
the scatter of the differences in the bins.  Panel $c$ shows the
number of points  (objects)  
in the bins.  Figure~\ref{figure:dline-ohr} shows that the difference between the
measured emission-line intensities in the SDSS-A and SDSS-B catalogues
changes with oxygen abundance.  The R$_{3}$ line intensities are
rather close in both catalogues. The N$_{2}$, H$\alpha$, and S$_{2}$
line intensities are higher in the SDSS-B catalogue at low metallicities
and the differences $F_{SDSS-B} - F_{SDSS-A}$ decrease with
metallicity and reach zero at the metallicity 12 + log(O/H) $\sim$
8.25.  The intensities of those lines in the SDSS-B catalogue are lower,
on average, than in the SDSS-A catalogue at high metallicities, 12 +
log(O/H) $\ga$ 8.25.  The absolute difference increases with abundance
and becomes as large as $\sim 0.1$ dex at metallicities of 12 +
log(O/H) $\ga$ 8.6.  The intensity of the R$_{2}$ line is larger in
the SDSS-B catalogue than in the SDSS-A catalogue at any metallicity and
the difference exceeds 0.1 dex at 12 + log(O/H) $\ga$ 8.6.

Figure~\ref{figure:ohmpa-doh} shows the comparison of the oxygen
abundances determined through the different calibrations using the
emission-line measurements from the SDSS-A and SDSS-B catalogues.  The
symbols represent the mean values of the abundance differences
for objects in bins
of 0.1 dex in (O/H)$_{SDSS-A}$. The centres of the bins for abundances
based on different calibrations are shifted in order to avoid the
coincidence of the positions of the symbols in the Figure.  The bars
denote the mean values of the scatter of the abundance differences in
the bins. 

Figure \ref{figure:ohmpa-doh} shows that the (O/H)$_{R}$
(as well as (O/H)$_{S}$ and (O/H)$_{N}$) abundances determined using
the emission-line measurements from the SDSS-A and SDSS-B catalogues
agree within around 0.02 dex for the objects with metallicities
12+log(O/H) $\ga$ 8.1.  The difference is larger at lower
metallicities and is $\sim$0.05 dex at 12+log(O/H) $\la$ 8.0.  The
difference between the (O/H)$_{P}$ abundances determined using the
emission-line measurements from the SDSS-A and SDSS-B catalogues is
large for the objects with metallicities 12+log(O/H) $\ga$ 8.55.  Here
the absolute value of the difference exceeds 0.1 dex.

Thus, the uncertainties in abundances estimated through the different
calibrations due to the uncertainties in the data reduction and line
flux measurements are within $\sim$0.02 dex for objects with
metallicities from 12 + log(O/H) $\sim$ 8.1 to 12 + log(O/H) $\sim$
8.55 and are higher at lower and higher metallicities.

It may appear surprising that considerably different line intensities
in the SDSS-A and SDSS-B catalogues (Fig.~\ref{figure:dline-ohr}) yield
very similar abundances (Fig.~\ref{figure:ohmpa-doh}). This can be
explained by the following.  Since the H${\alpha}$/H${\beta}$ ratio is
higher in the SDSS-A catalogue and, consequently, the extinction
parameter $C_{{\rm H}{\beta}}$ is also higher, the dereddening results
in a decrease of the systematic difference in the measured line
intensities between the SDSS-A and SDSS-B catalogues.  We have also
dereddened the SDSS-A and SDSS-B spectra using the reddening function
from \citet{Cardelli1989} for $R_{V}$ = 3.1 to check the robustness of
the results. We assume $C_{{\rm H}{\beta}} = 0.47A_{V}$ following
\citet{Lee2005}.  There is no significant difference between the
results obtained with Whitford's and Cardelli et al.'s reddening laws.

It is thus difficult to choose which catalogue (SDSS-A or SDSS-B) is
more reliable.

\subsection{Abundances determined from H\,{\sc ii} region slit spectra,
  from SDSS fibre spectra, and from MaNGA spaxel spectra}

\begin{figure}
\resizebox{1.00\hsize}{!}{\includegraphics[angle=000]{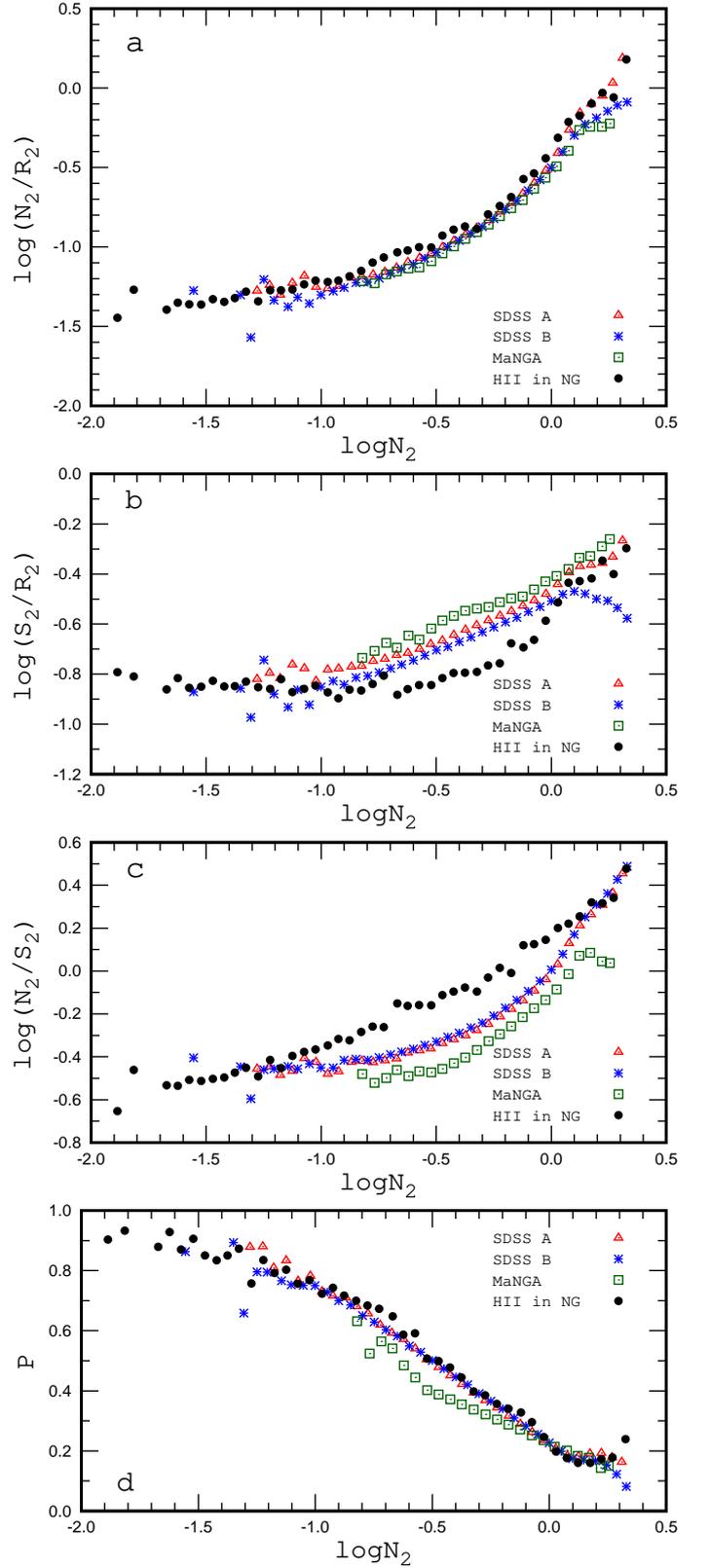}}
\caption{Panel $a$ shows the nitrogen to oxygen line ratio N$_{2}$/R$_{2}$  
as a function of the  N$_{2}$  line intensity for the SDSS fibre 
spectra (catalogues SDSS-A and SDSS-B), for the spaxel MaNGA spectra, 
and for the slit spectra of H\,{\sc ii} regions in nearby galaxies. 
The symbols represent the mean values for objects in bins of 0.05 dex in logN$_{2}$. 
Panels $b, c$, and  $d$ show the same but for the sulphur to oxygen 
line ratio S$_{2}$/R$_{2}$, for the nitrogen to sulphur 
line ratio N$_{2}$/S$_{2}$, and for the excitation parameter $P$, respectively.
}
\label{figure:ln2-lx}
\end{figure}

\begin{figure}
\begin{center}
\resizebox{1.00\hsize}{!}{\includegraphics[angle=000]{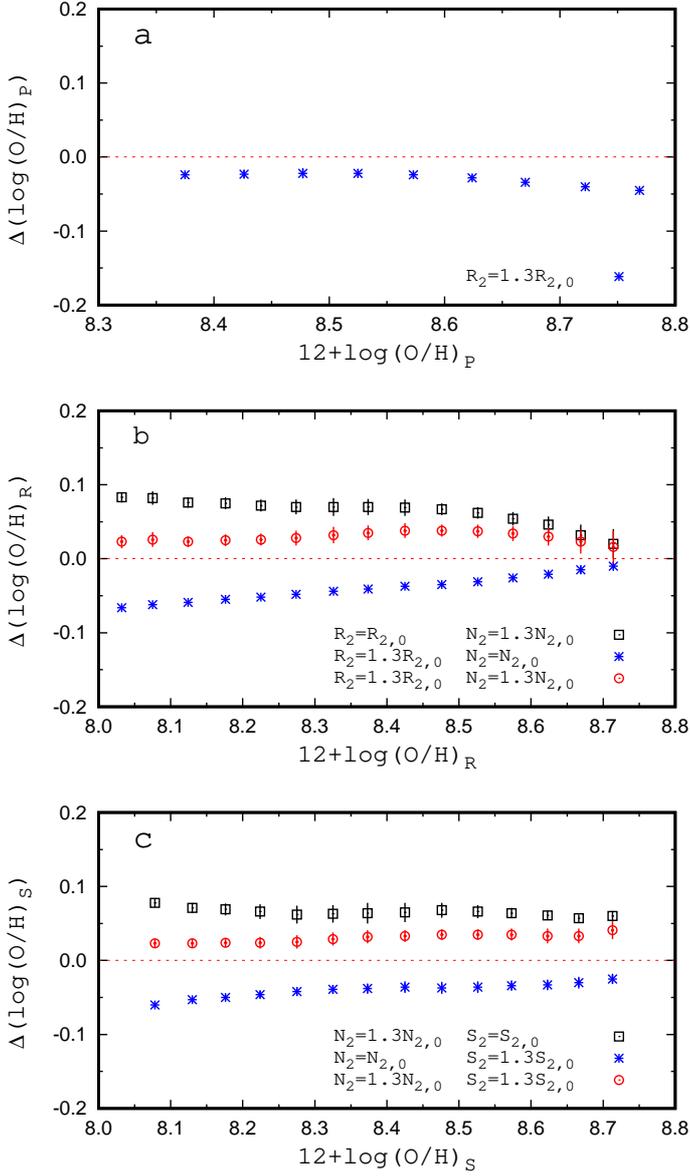}}
\caption{
Influence of the increase of the emission-line fluxes on the oxygen
abundances obtained through different calibrations.  Panel $a$. The
difference between oxygen abundances in H\,{\sc ii} regions in nearby
galaxies obtained with the original R$_{2}$=R$_{2,0}$ emission line
and with the emission line R$_{2}$=1.3R$_{2,0}$ increased by a factor
of 1.3.  The symbols represent the mean values of the abundance
differences for objects in bins of 0.05 dex in (O/H). The bars (comparable with
the symbol size) denote the mean values of the scatter of the
differences in the bins. 
Panel $b$ shows the influence of the increase of the R$_{2}$ and
N$_{2}$ emission-line fluxes on the oxygen abundances (O/H)$_{R}$.
Panel $c$ shows the influence of the increase of the N$_{2}$ and
S$_{2}$ emission-line fluxes on the oxygen abundances (O/H)$_{S}$.
}
\label{figure:ohx-dohx-plus}
\end{center}
\end{figure}

\begin{figure}
\begin{center}
\resizebox{0.92\hsize}{!}{\includegraphics[angle=000]{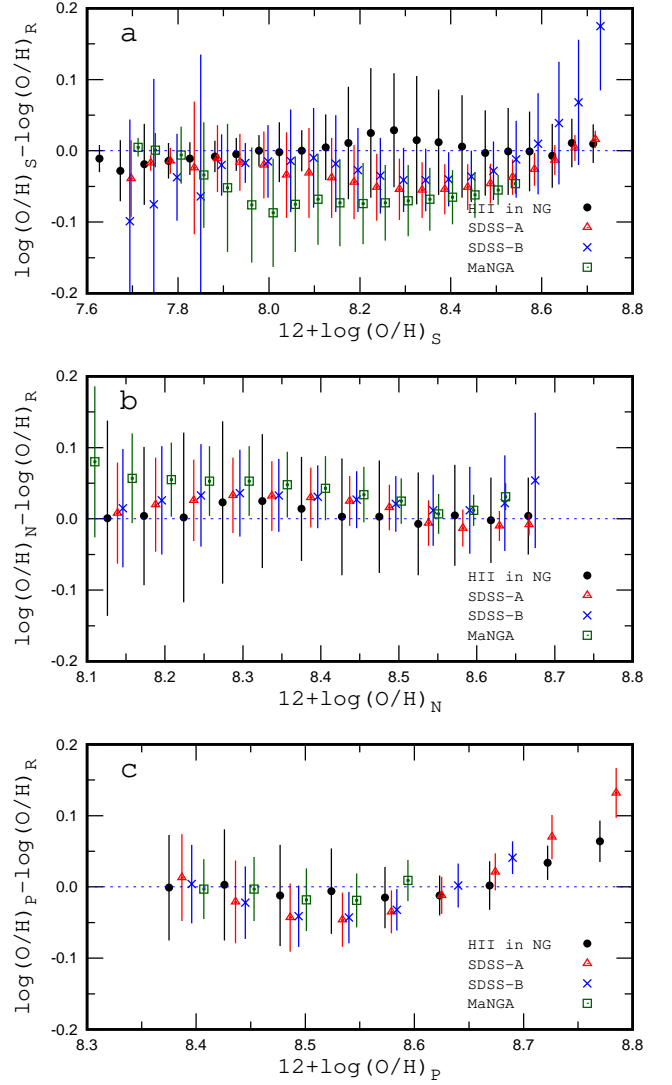}}
\caption{
The comparison between oxygen abundances determined through different
calibrations based on the slit spectra of H\,{\sc ii} regions in
nearby galaxies, the fibre SDSS spectra (with lines from the SDSS-A
and SDSS-B catalogues), and the MaNGA spaxel spectra.  The symbols
represent the mean values of the differences for objects in bins of 0.05 dex in
O/H, and the bars show the scatter of differences in these bins. 
}
\label{figure:oh-doh-samples}
\end{center}
\end{figure}

\begin{figure*}
\begin{center}
\resizebox{1.00\hsize}{!}{\includegraphics[angle=000]{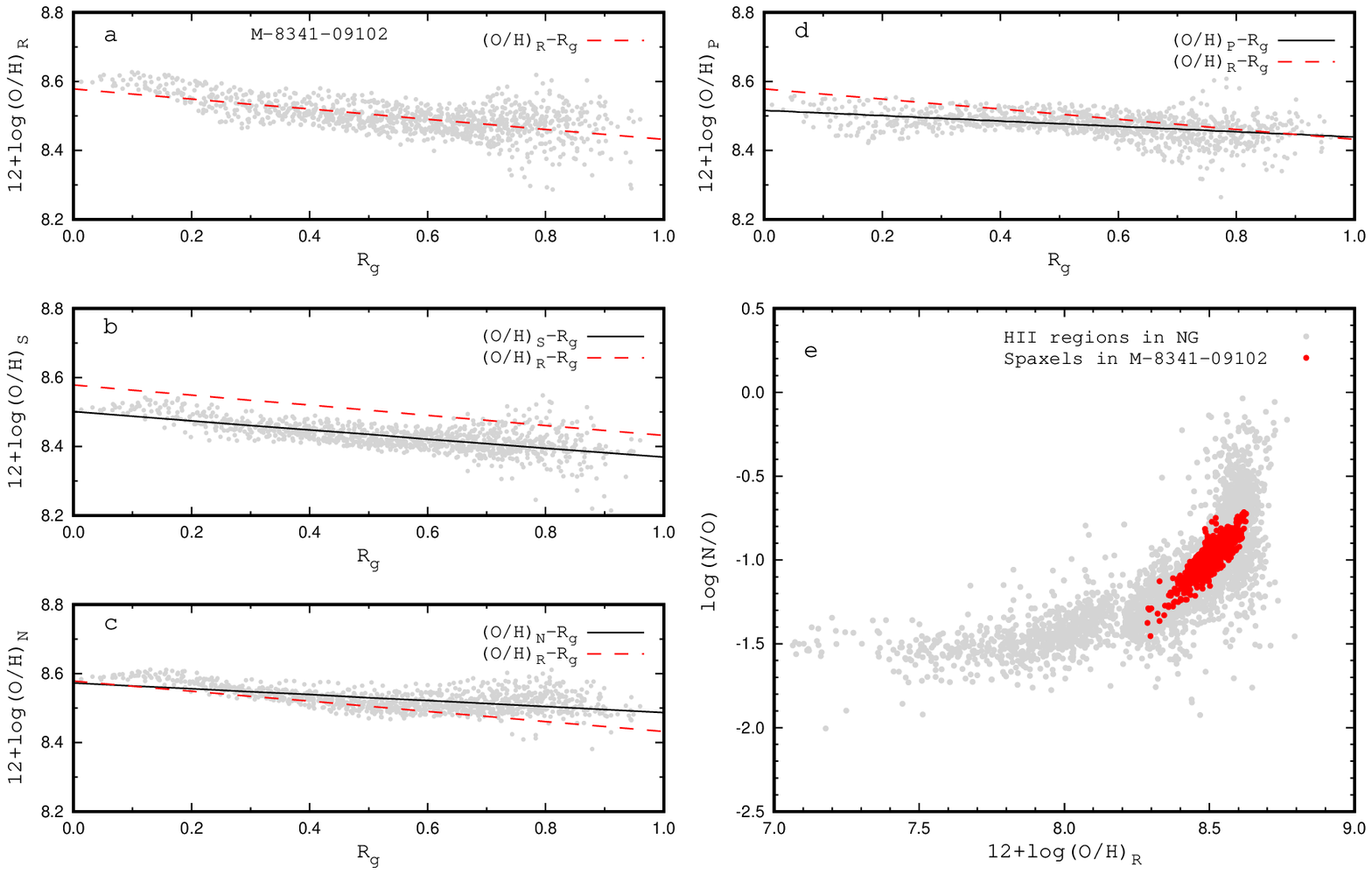}}
\caption{
Panels $a, b, c$, and $d$ show radial distributions of the oxygen abundances
determined through the different calibrations in the disc of the MaNGA
galaxy M-8341-09102.  The grey points are abundances in the individual
spaxels.  The line is the best fit to those data. The dashed line in
each panel is the $R_g$ -- (O/H)$_{R}$ relation.  Panel $e$ shows the
(O/H)$_{R}$ -- N/O diagram. The grey points are abundances in the
H\,{\sc ii} regions in the nearby galaxies; the red points are
abundances in the individual spaxels of the MaNGA galaxy M-8341-09102. 
}
\label{figure:ohx-rg-m-8341-09102}
\end{center}
\end{figure*}

Here we compare the abundances determined through different
calibrations from the H\,{\sc ii} region slit spectra, from the SDSS
fibre spectra, and from the MaNGA spaxel spectra.

\subsubsection{Line ratios}

Firstly we compare the line intensities in the spectra of our 
samples.  The objects of each sample are binned in the  N$_{2}$ line
intensities. The  N$_{2}$ value is an indicator of the electron
temperature and abundance in a H\,{\sc ii} region; for an example, see Fig. 2
in \citet{Pilyugin2016}. Thus the binning in the   N$_{2}$ values can
be considered as a binning in abundances.  Furthermore, the binned
N$_{2}$, R$_{2}$, and S$_{2}$ line intensities show a large scatter
that can mask any differences in line intensities between different
samples of objects. Therefore it is preferable to compare not the line
intensities but the ratio of the line intensities. 

Panel $a$ of Fig.~\ref{figure:ln2-lx} shows the  N$_{2}$/R$_{2}$ line
ratio as a function of  N$_{2}$ line intensity for our samples of slit
spectra of H\,{\sc ii} regions in nearby galaxies, for the SDSS fibre
spectra (catalogues SDSS-A and SDSS-B), and for the MaNGA spaxel
spectra.  The symbols represent the mean values in bins of 0.05 dex in
N$_{2}$.  The increase of N$_{2}$ or equal increases of N$_{2}$ and
R$_{2}$ would result in a shift of the points along the X-axis towards
a larger N$_{2}$ value.  The increase of R$_{2}$ or different
increases of R$_{2}$ and  N$_{2}$ would result in a shift of the
points along the Y-axis.  Inspection of Fig.~\ref{figure:ln2-lx} shows
that the  N$_{2}$/R$_{2}$ -- N$_{2}$ diagrams for the SDSS and the
MaNGA objects are in agreement with each other but there is some
difference to those for the H\,{\sc ii} regions in nearby galaxies in
the sense that there is shift of the N$_{2}$/R$_{2}$ -- N$_{2}$
diagrams for the SDSS and the MaNGA objects relative to the
N$_{2}$/R$_{2}$ -- N$_{2}$ diagram for the slit spectra of $\sim 0.1$
dex along the X- or Y-axis.

It has been known for a long time that there is no one-to-one
correspondence between the oxygen and nitrogen abundances in galaxies
since there is a time delay in the enrichment of the interstellar
medium in nitrogen as compared to oxygen. The nitrogen-to-oxygen ratio
at a given oxygen abundance depends on the star formation history in a
galaxy, that is, N/O is an indicator of the time that has
elapsed since the last episode of star formation \citep{Edmunds1978}.
Thus, the difference in the N$_{2}$/R$_{2}$ -- N$_{2}$ diagrams may
seem to suggest that there are differences in the star formation
histories of the sample of nearby galaxies and in  the galaxy samples
between the SDSS and MaNGA surveys.  On the other hand, the difference in the
N$_{2}$/R$_{2}$ -- N$_{2}$ diagrams may suggest that the strengths of
the N$_{2}$ and/or R$_{2}$ in the spaxel (fibre) spectra are increased
by the contribution of the radiation of the diffuse ionized gas.  If
this is the case then the increase of the strengths of the N$_{2}$
and/or R$_{2}$ does not exceed a factor of $\sim 1.3$.

Panels $b$ and $c$ of Fig.~\ref{figure:ln2-lx} show the
S$_{2}$/R$_{2}$ and N$_{2}$/S$_{2}$ line ratios as a function of
N$_{2}$  for those samples of objects.  Inspection of these two panels
shows that the S$_{2}$/R$_{2}$ -- N$_{2}$ and N$_{2}$/S$_{2}$ --
N$_{2}$ diagrams for the MaNGA galaxies differ from the ones for the
SDSS galaxies and for the nearby galaxies. The difference is maximized
between the S$_{2}$/R$_{2}$ -- N$_{2}$ (and N$_{2}$/S$_{2}$ --
N$_{2}$) diagrams for the MaNGA and the nearby galaxies.  The
difference changes with N$_{2}$. The difference in S$_{2}$/R$_{2}$ and
N$_{2}$/S$_{2}$ is around 0.3 dex at logN$_{2} \sim -0.5$.  If the
contribution of the radiation of the diffuse ionized gas to the spaxel
(fibre) spectra is responsible for the differences in the
S$_{2}$/R$_{2}$ and N$_{2}$/S$_{2}$ line ratios then this shows that
the enhancement of the strength of S$_{2}$ due to the contribution of
the diffuse ionized gas amounts to up to a factor of two. 

Panel $d$ of Fig.~\ref{figure:ln2-lx} shows the excitation parameter
$P$ as a function of N$_{2}$ for those samples of objects.  Again, the
largest difference between the excitation parameters at a given
N$_{2}$ occurs between the MaNGA spaxel spectra and  H\,{\sc ii}
regions in nearby galaxies.

\subsubsection{Influence of the line strength increase on the obtained abundances}

Before comparing the abundances determined through different
calibrations from the H\,{\sc ii} region slit spectra, the SDSS fibre
spectra, and the MaNGA spaxel spectra, we examine the influence of the
increase of the emission-line strengths on the oxygen abundances
obtained through different calibrations. This can be examined by
comparing oxygen abundances in H\,{\sc ii} regions in nearby galaxies
obtained with the original strengths of the emission lines and with
the increased strengths.  The artificial increase of the emission-line
intensities can be considered as a rough simulation of the
contribution of the radiation of the diffuse ionized gas to the
composite spectra.

It is evident that the change of the value of the (O/H)$_{N}$
abundance depends on the change of N$_{2}$, but does not depend on the
change of R$_{2}$ and S$_{2}$.  It can easily be seen from
Eq.~\ref{equation:ohn} that the increase of the strength of the
N$_{2}$ by a factor of 1.3 results in an increase of (O/H)$_{N}$ by
0.06 dex.  It turn, the change of the value of the (O/H)$_{P}$
abundance depends on the change of R$_{2}$, but does not depend on the
change of N$_{2}$ and S$_{2}$.  Panel $a$ of
Fig.~\ref{figure:ohx-dohx-plus} shows the difference between the
oxygen abundances (O/H)$_{P}$ in H\,{\sc ii} regions in nearby
galaxies obtained with the original emission line R$_{2}=$ R$_{2,0}$
and with the emission line R$_{2}=1.3$ R$_{2,0}$ increased by a factor
of 1.3 in each individual spectrum.  The symbols represent the mean
values of the abundance differences in bins of 0.05 dex in
(O/H)$_{P}$. The bars (comparable with the symbol size) denote the
mean values of the scatter of the differences in the bins.  Panel $a$
of Fig.~\ref{figure:ohx-dohx-plus} shows that the increase of the
R$_{2}$ emission-line intensity by a factor of 1.3 results in a
decrease of the (O/H)$_{P}$ abundance by around 0.03 dex. 

Panel $b$ of Fig.~\ref{figure:ohx-dohx-plus}  shows the influence of
the increase of the R$_{2}$ and N$_{2}$ emission-line strengths on the
(O/H)$_{R}$ abundance.  One can see that an increase of the R$_{2}$
emission-line intensity by a factor of 1.3 leads to a decrease of the
(O/H)$_{R}$ abundance by around 0.06 dex at 12+log(O/H) $\sim 8.1$ and
by around 0.015 dex at 12+log(O/H) $\sim 8.65$.  Increasing the
N$_{2}$ emission-line intensity by a factor of 1.3 results in an
increase of the (O/H)$_{R}$ abundance by around 0.07 dex at
12+log(O/H) $\sim 8.1$ and by around 0.04 dex at 12+log(O/H) $\sim
8.65$.  An increase of the  R$_{2}$ line intensity results in the
opposite effect on the (O/H)$_{R}$ abundance as compared to the
increase of the N$_{2}$ line intensity. Thus, an increase of both the
R$_{2}$ and N$_{2}$ line intensities by a factor of 1.3 results in an
increase of the (O/H)$_{R}$ abundance of $\sim 0.04$ dex.

Panel $c$ of Fig.~\ref{figure:ohx-dohx-plus} shows the influence of
the increase of the N$_{2}$ and S$_{2}$ emission-line strengths on the
(O/H)$_{S}$ abundance.  Again, an increase of the  S$_{2}$ line
intensity results in the opposite effect on the (O/H)$_{S}$ abundance
as compared to the increase of the N$_{2}$ line intensity, and the
increase of both the S$_{2}$ and N$_{2}$ line intensities by a factor
of 1.3 leads to an increase of the (O/H)$_{S}$ abundance by $\sim
0.04$ dex.

Thus, an increase of the R$_{2}$, N$_{2}$, and S$_{2}$ line strengths
by a factor of 1.3 results in an increase of the (O/H)$_{N}$ abundance
by as much as 0.06 dex, a decrease of the (O/H)$_{P}$ abundance of
around 0.03 dex, and an increase of the (O/H)$_{R}$ and (O/H)$_{S}$
abundances by $\sim 0.04$ dex.

\subsubsection{Comparison between abundances determined through different calibration
  in our samples of objects}

Panel $a$ of Fig.~\ref{figure:oh-doh-samples} shows the differences
between the (O/H)$_{S}$ and (O/H)$_{R}$ oxygen abundances determined
from the slit spectra of H\,{\sc ii} regions in nearby galaxies, from
the SDSS fibre spectra (for the SDSS-A and SDSS-B catalogues), and from
the MaNGA spaxel spectra.  The symbols represent the mean values of
the differences for objects  
in bins of 0.05 dex in (O/H)$_{S}$. The bars denote
the mean values of the scatter of the differences
 for objects in the bins. 
Examination of Panel $a$ of Fig.~\ref{figure:oh-doh-samples} shows
that the difference between the (O/H)$_{S}$ and (O/H)$_{R}$ oxygen
abundances determined from the slit spectra of H\,{\sc ii} regions in
nearby galaxies is rather small, usually lower than 0.02 dex.

The (O/H)$_{R}$ -- (O/H)$_{S}$ diagrams for the SDSS objects obtained
with the lines from the SDSS-A and SDSS-B catalogues are similar for
metallicities from 12 + log(O/H)$_{S} \sim 7.9$ to  12 + log(O/H)$_{S}
\sim 8.6$ and they differ significantly at lower (12 + log(O/H)$_{S}
\la 7.9$) and higher (12 + log(O/H)$_{S} \ga 8.6$) metallicities.
Thus, the comparison between (O/H)$_{R}$ and (O/H)$_{S}$ abundances in
the SDSS objects is justified for objects with metallicities from 12
+ log(O/H)$_{S}$ $\sim$ 7.9 to  12 + log(O/H)$_{S}$ $\sim$ 8.6 only.
The (O/H)$_{S}$ abundances are lower than the (O/H)$_{R}$ abundances,
and the difference is as large as $\sim 0.05$ dex for objects with
metallicities from 12 + log(O/H) $\sim 8.2$ to 12 + log(O/H) $\sim
8.5$. The (O/H)$_{S}$ abundances in the MaNGA objects are also lower
than the (O/H)$_{R}$ abundances, and the difference is larger (up to
around 0.1 dex) than for the SDSS objects. This is not surprising
since the S$_{2}$ line intensity is significantly enhanced in the SDSS
fibre and MaNGA spaxel spectra.

Panel $b$ of Fig.~\ref{figure:oh-doh-samples} shows the differences
between the (O/H)$_{N}$ and (O/H)$_{R}$ abundances for our samples of
objects. The applicability of our $N$ calibration is restricted by the
condition logN$_{2} \ga -0.6$, hence the  (O/H)$_{N}$ abundances in
low-metallicity objects cannot be determined. One can see that the
difference between the (O/H)$_{N}$ and (O/H)$_{R}$ abundances for the
H\,{\sc ii} regions in nearby galaxies is small, usually lower than
0.01 dex, except for objects with abundances of 12 + log(O/H) $\sim
8.3$ where the difference is around 0.025 dex.  Again, the (O/H)$_{R}$
-- (O/H)$_{N}$ diagrams for the SDSS objects obtained with lines from
the SDSS-A and SDSS-B catalogues are similar to each other for
metallicities 12 + log(O/H)$_{N} \la 8.6$, but differ for higher
metallicities. The (O/H)$_{N}$ abundances are higher than the
(O/H)$_{R}$ abundances, and the difference is within 0.03 dex, which
is comparable to the maximum difference between the (O/H)$_{N}$ and
(O/H)$_{R}$ abundances for the H\,{\sc ii} regions in nearby galaxies.
The (O/H)$_{N}$ abundances in the MaNGA objects are also higher than
the (O/H)$_{R}$ abundances. The difference increases with decreasing
metallicity and is around 0.05 dex for objects with 12 + log(O/H)$_{N}
\la 8.3$.

Close examination of Fig.~\ref{figure:oh-doh-samples} and
Fig.~\ref{figure:ln2-lx} suggests that the discrepancy between the
(O/H)$_{N}$ and (O/H)$_{R}$ abundances for the MaNGA galaxies can be
explained by the intrinsic properties of the N$_{2}$ calibration.
Indeed, the $N$ calibration can produce an incorrect (O/H)$_{N}$ even if
the strength of N$_{2}$ is correct. Like any other 1D
calibration, the N$_{2}$ calibration encounters the following problem.
The  N$_{2}$ line intensity in the spectrum of H\,{\sc ii} region
depends not only on the oxygen abundance (O/H) but also on its
excitation $P$ and the nitrogen-to-oxygen abundance ratio N/O, that is,
N$_{2}$ = $f$(O/H,$P$,N/O) or  (O/H)$_{N}$ = $f$(N$_{2}$,$P$,N/O).
The calibrating data points occupy a wide band in the O/H -- N$_{2}$
diagram, that is, the scatter around the 1D calibration
relation O/H = $f$(N$_{2}$) is large \citep{Pettini2004, Marino2013}.
The deviation of the calibrating   H\,{\sc ii} region from the O/H =
$f$(N$_{2}$) relation  depends on its excitation and the N/O abundance
ratio. Of course, the uncertainties in the oxygen abundances in the
calibrating data points also contribute to the scatter around the O/H
= $f$(N$_{2}$) relation.  The calibration relation (O/H)$_{N}$ =
$f$(N$_{2}$) produces the correct (O/H)$_{N}$ abundance in a target
region (spaxel) if only its excitation $P$ and nitrogen-to-oxygen
abundance ratio N/O correspond to those parameters in the calibrating
points (with a given value of N$_{2}$) located close to the (O/H) =
$f$(N$_{2}$) relation  in the O/H  -- N$_{2}$ diagram.  If the
excitation $P$ and/or the nitrogen-to-oxygen abundance ratio N/O in
the target spaxel region differs from the those values in the
calibrating points located close to the O/H = $f$(N$_{2}$) relation
then the N$_{2}$ calibration produces incorrect (O/H)$_{N}$ abundance.
Then the disagreement between the (O/H)$_{N}$ and (O/H)$_{R}$
abundances for the MaNGA galaxies (Panel $b$ of
Fig.~\ref{figure:oh-doh-samples}) can be explained by incorrect
(O/H)$_{N}$ abundances caused by the systematic difference in the
excitations $P$ of the MaNGA spaxel spectra and H\,{\sc ii} regions in
nearby galaxies (Panel $d$ of Fig.~\ref{figure:ln2-lx}) used as the
calibrating data points in deriving the O/H  -- N$_{2}$ relation.
This expectation will be checked below through the comparison between
the radial distributions of the (O/H)$_{N}$ and (O/H)$_{R}$ abundances
in the MaNGA galaxies.

Panel $c$ of Fig.~\ref{figure:oh-doh-samples} shows the differences
between the (O/H)$_{P}$ and (O/H)$_{R}$ abundances for our samples of
objects. The applicability of the upper-branch $P$ calibration is
restricted by the condition 12 + log(O/H) $\ga$ 8.35. Therefore the
(O/H)$_{P}$ abundances are determined for high-metallicity objects
only.  On the other hand, the (O/H)$_{P}$ abundances at high
metallicities (12 + log(O/H) $\ga 8.55$) can involve a large error due
to the uncertainty in the data reduction and emission line
measurements; see Fig.~\ref{figure:ohmpa-doh}.  Thus, the (O/H)$_{P}$
abundances are more or less reliable in a narrow interval of
metallicities only.  Therefore, a reliable distribution of the
(O/H)$_{P}$ abundances can be found only in a small number of
galaxies. 

It should be noted that the difference between abundances estimated
through different calibrations for individual objects can be
significantly larger than the mean value of the differences.  The mean
scatter of the differences is shown by the error bars in
Fig.~\ref{figure:oh-doh-samples}.

\section{Abundance distributions in the MaNGA galaxies}

\begin{figure}
\begin{center}
\resizebox{1.00\hsize}{!}{\includegraphics[angle=000]{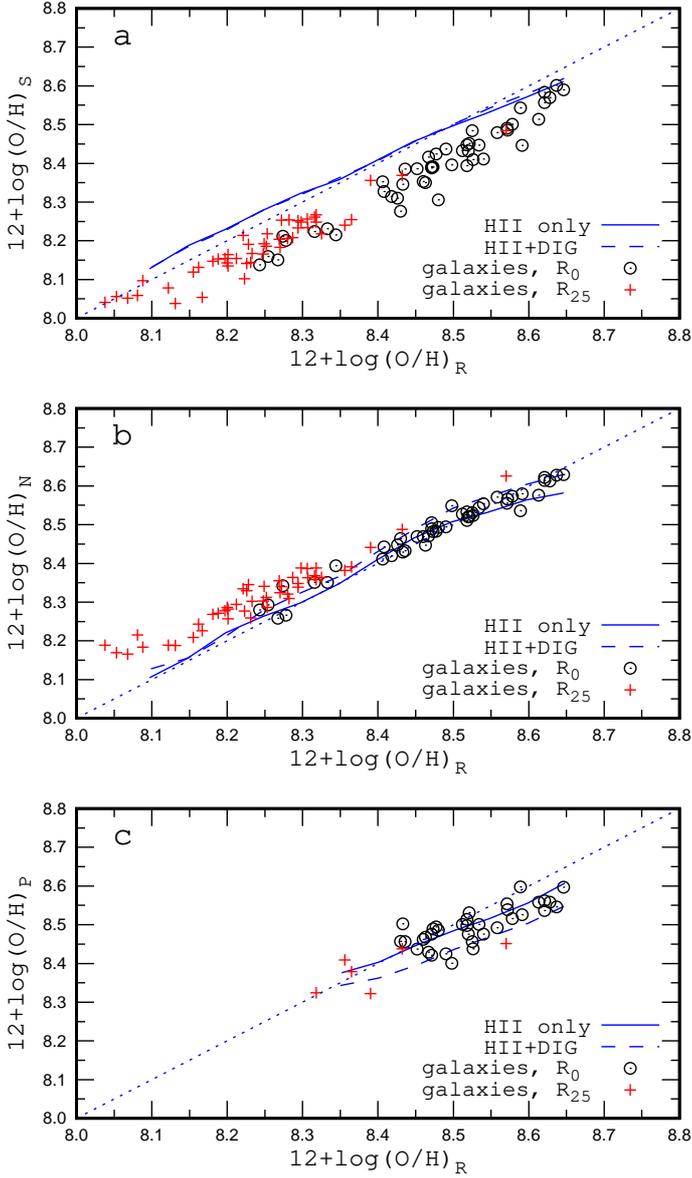}}
\caption{
Comparison of the (O/H)$_{R}$ oxygen abundances with abundances
estimated through other calibrations for our sample of MaNGA galaxies.
The circles in each panel show the central (intersect) oxygen
abundances. The plus symbols indicate abundances (intersect values) at
the optical radius $R_{25}$.  The solid line shows the oxygen
abundances for the sample of H\,{\sc ii} regions in the nearby
galaxies.  The long-dashed line shows the oxygen abundances for those
H\,{\sc ii} regions obtained with the lines $R_{2}$,  $N_{2}$,  and
$S_{2}$ increased by a factor of 1.3 to simulate the DIG contribution
to the spectra.  The short-dashed line indicates unity.   
}
\label{figure:ohr-ohx-manga}
\end{center}
\end{figure}

\begin{figure*}
\begin{center}
\resizebox{1.00\hsize}{!}{\includegraphics[angle=000]{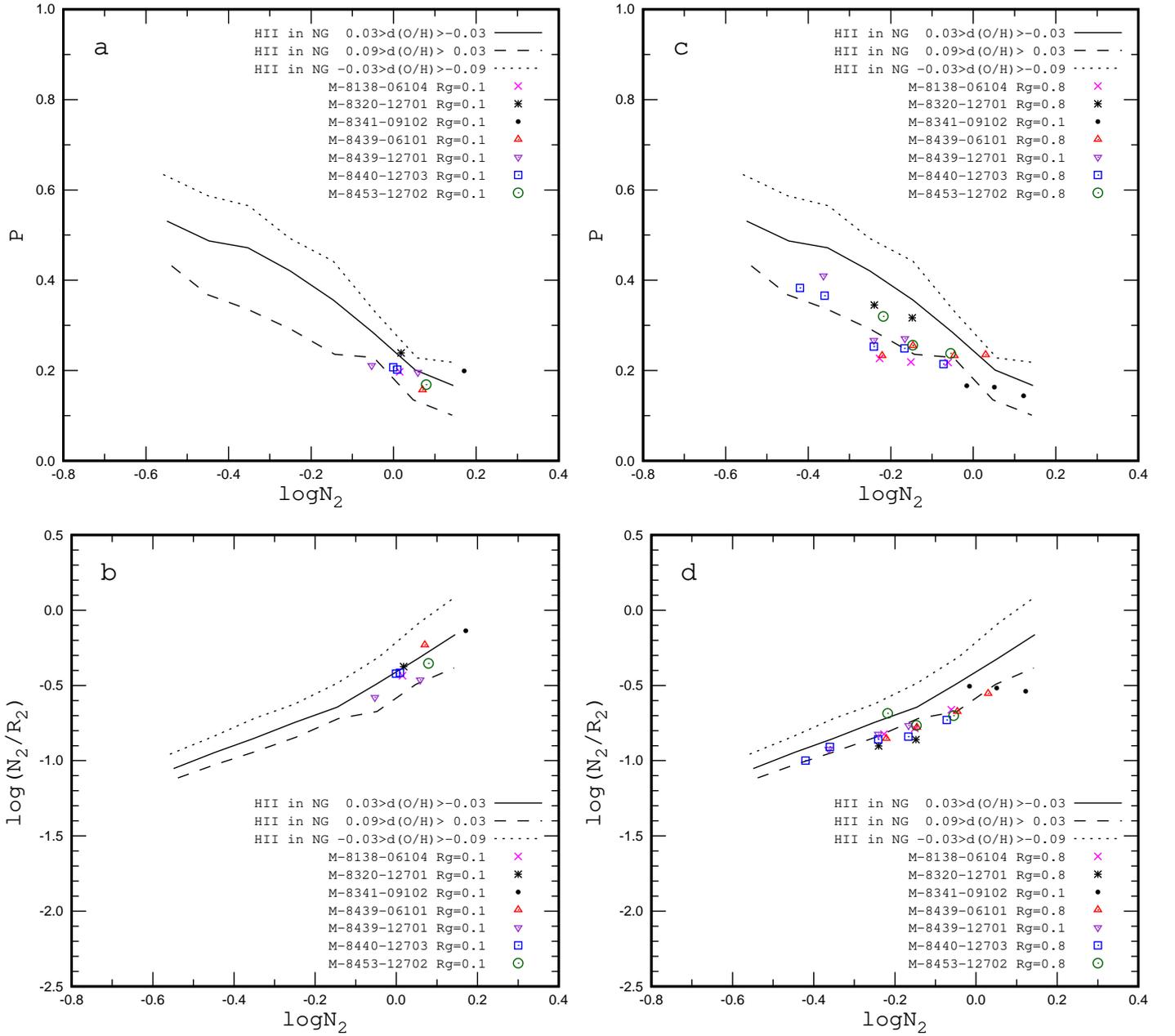}}
\caption{
The symbols in Panel $a$ show the excitation parameter $P$ as a
function of the $N_{2}$ line intensity for the spaxels in central
regions ($R_{g} < 0.1$) of several MaNGA galaxies.  The mean values of
$P$ for objects in bins of 0.1 dex in logN$_{2}$ are presented.  The solid line
shows the excitation parameter $P$ as a function of $N_{2}$ for the
H\,{\sc ii} regions in the nearby galaxies with $0.03 >$
log(O/H)$_{N}$ -- log(O/H)$_{R} > -0.03$.  The long-dashed line shows
$P$  for H\,{\sc ii} regions with $0.09 >$ log(O/H)$_{N}$ --
log(O/H)$_{R} > 0.03$, and the short-dashed line shows $P$ for H\,{\sc ii} regions
with $-0.03 >$ log(O/H)$_{N}$ -- log(O/H)$_{R} > -0.09$.  Panel $b$
shows the same as Panel $a$ but for the N$_{2}$/R$_{2}$ line ratio.
Panel $c$ shows the same as Panel $a$ but for spaxels with
galactocentric distances from 0.775 to 0.825 $R_{g}$.  Panel $d$ shows
the same as Panel $b$ but for spaxels with galactocentric distances
from 0.775 to 0.825 $R_{g}$.
}
\label{figure:ln2-p-manga}
\end{center}
\end{figure*}

\begin{figure}
\begin{center}
\resizebox{1.00\hsize}{!}{\includegraphics[angle=000]{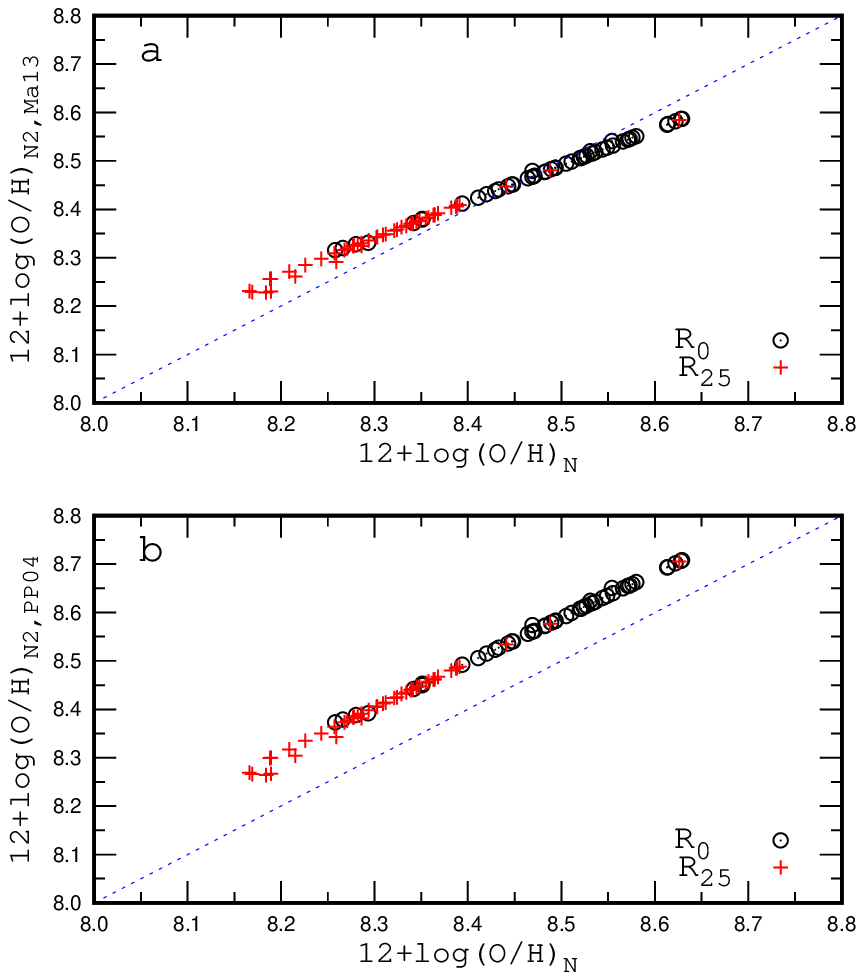}}
\caption{
Comparison of the oxygen abundances estimated through different N2
calibration relations.  The upper panel shows the (O/H)$_{N2,Ma13}$
abundances obtained through the N2 calibration relation of
\citet{Marino2013} versus\ the (O/H)$_{N}$ abundances determined using
our expression for the N calibration.  The circles indicate the
central (intersect) oxygen abundances for our sample of MaNGA
galaxies. The plus symbols show abundances (intersect values) at the
optical radius $R_{25}$.  The line is that of equal values.  The lower
panel shows the same as the upper panel but for the (O/H)$_{N2,PP04}$
abundances obtained through the N2 calibration relation of
\citet{Pettini2004}.
}
\label{figure:ohn-ohman-ohppn}
\end{center}
\end{figure}

The (O/H)$_{R}$, (O/H)$_{S}$, (O/H)$_{N}$, and (O/H)$_{R}$ abundances
were estimated from the spaxel spectra in the discs of the MaNGA
galaxies of our sample. The radial abundance distributions were fitted
by linear relations. The obtained (O/H) -- $R_{g}$ relation produces
the central (intersect) abundance and abundance at the optical radius
$R_{25}$ (also the intersect value). Since the $P$ calibration is
applicable to high-metallicity objects only, the (O/H)$_{P}$ --
$R_{g}$ relation was determined for spaxels with galactocentric
distances of less than $R^{*}$, where $R^{*}$ is the radius at which the
12+log(O/H)$_{R} = 8.4$. If $R^{*} < 0.3 R_{25}$ then the (O/H)$_{P}$
-- $R_{g}$ relation was not determined and the central abundance could
not be estimated.  If $0.3R_{25} < R^{*} < 0.7R_{25}$ then the
(O/H)$_{P}$ -- $R_{g}$ relation was used to estimate the central
abundance but the abundance at the optical radius was not measured.
If $R^{*} > 0.7R_{25}$ then both the central abundance and the
abundance at the optical radius were estimated from the (O/H)$_{P}$ --
$R_{g}$ relation.

The (O/H)$_{R}$ -- N/O diagram was also constructed for each MaNGA
galaxy and compared to the diagram for the H\,{\sc ii} regions in
nearby galaxies.  The N/O abundances ratio was estimated using the
N/O calibration relation from \citet{Pilyugin2016}.  This diagram
serves for visual control of the validity of the obtained abundances.

Panel $a$ of Fig.~\ref{figure:ohx-rg-m-8341-09102} shows the radial
distribution of the (O/H)$_{R}$ abundances across the disc of the
MaNGA galaxy M-8341-09102.  The grey points are abundances in the
individual spaxels.  The dashed line is the best fit to those data.
Panels $b, c,$ and $d$ of Fig.~\ref{figure:ohx-rg-m-8341-09102} show
the radial distributions of the (O/H)$_{S}$, (O/H)$_{N}$, and
(O/H)$_{P}$ abundances in the disc of M-8341-09102, respectively. The
grey points are abundances in the individual spaxels, and the solid
line depicts the best fit to those data.  The dashed line in Panels
$b, c$, and $d$ is the $R_g$ -- (O/H)$_{R}$ relation from the Panel
$a$.  Panel $e$ shows the (O/H)$_{R}$ -- N/O diagram. The grey points
denote abundances in the H\,{\sc ii} regions in the nearby galaxies;
the red points are abundances in the individual spaxels of the MaNGA
galaxy M-8341-09102. 

Panel $a$ of Fig.~\ref{figure:ohr-ohx-manga} shows the comparison
between the (O/H)$_{s}$ and (O/H)$_{R}$ abundances for our sample of
MaNGA galaxies.  The circles mark the central (intersect) oxygen
abundances. The plus signs mark the abundances (intersect values) at
the optical radius $R_{25}$.  The short-dashed line is that of equal
values. For comparison, the solid line shows the
(O/H)$_{s}$ versus (O/H)$_{R}$ abundances (mean values in bins of 0.05
dex in (O/H)$_{R}$) for our sample of H\,{\sc ii} regions in the
nearby galaxies).  The long-dashed line shows the (O/H)$_{s}$ versus
(O/H)$_{R}$ abundances for those H\,{\sc ii} regions obtained with the
lines R$_{2}$,  N$_{2}$,  and S$_{2}$ increased by a factor of 1.3 to
simulate the DIG contribution to the spectra.

Panel $a$ of Fig.~\ref{figure:ohr-ohx-manga} shows that the
(O/H)$_{S}$ abundances are lower than the (O/H)$_{R}$ abundances both
in the centres and at the optical radii of the MaNGA galaxies. The
increase of the lines  N$_{2}$ and S$_{2}$ by a factor of 1.3 cannot
explain the difference between the (O/H)$_{S}$ and (O/H)$_{R}$
abundances. This is not surprising. Indeed, Panels $b$ and $c$ of
Fig.~\ref{figure:ln2-lx} show that the sulphur line S$_{2}$ is
increased in the MaNGA spectra by a factor up to $\sim$2. 

Panel $b$ of Fig.~\ref{figure:ohr-ohx-manga} shows the comparison
between the (O/H)$_{N}$ and (O/H)$_{R}$ abundances for our sample of
MaNGA galaxies.  The circles stand for the central abundances; the
plus signs are the abundances at the optical radii $R_{25}$. Again,
the solid line shows (O/H)$_{N}$ as a function of the (O/H)$_{R}$
abundance for our sample of H\,{\sc ii} regions in nearby galaxies,
and the long-dashed line denotes the oxygen abundances for H\,{\sc ii}
regions obtained with the lines $R_{2}$ and $N_{2}$ increased by a
factor of 1.3 to simulate the DIG contribution to the spectra.

Panel $b$ of Fig.~\ref{figure:ohr-ohx-manga} shows that the central
(O/H)$_{N}$ and (O/H)$_{R}$ abundances in the MaNGA galaxies are close
to each other. They are compatible with the (O/H)$_{N}$ -- (O/H)$_{R}$
diagram for H\,{\sc ii} regions in nearby galaxies. The
increase of the lines $R_{2}$ and/or $N_{2}$ in the MaNGA spaxel
spectra is less than a factor of 1.3 (if any). This is strong
evidence showing that the central (O/H)$_{N}$ and (O/H)$_{R}$ abundances are
relatively reliable.

At the same time, Panel $b$ of Fig.~\ref{figure:ohr-ohx-manga} shows
that the (O/H)$_{N}$ abundances at the optical radii are
systematically higher than the (O/H)$_{R}$ abundances. We
suggest above that the discrepancy between the (O/H)$_{N}$ and
(O/H)$_{R}$ abundances determined from the MaNGA spectra can be
explained by the fact that the N-calibration produces incorrect abundances if
the excitation $P$ and/or the nitrogen-to-oxygen abundance ratio, N/O,
in an object differs from the values in the calibrating points
located in the (O/H)$_{R}$ -- N${_2}$ diagram close to the (O/H) =
$f$(N$_{2}$) relation.  We examine this expectation here.

Panel $a$ of Fig.~\ref{figure:ln2-p-manga} shows the excitation
parameter $P$ as a function of the $N_{2}$ line intensity for the
spaxels in the central regions ($R_{g} < 0.1$) of several MaNGA
galaxies.  The mean values of $P$ in bins of 0.1 dex in N$_{2}$ are
presented.  The solid line shows the excitation parameter $P$ as a
function of $N_{2}$ for our H\,{\sc ii} regions in nearby galaxies
with $0.03 >$ log(O/H)$_{N}$ -- log(O/H)$_{R} > -0.03$.  The H\,{\sc
ii} regions with correct (O/H)$_{N}$ abundances are located close to
the O/H = $f$(N$_{2}$) relation in the (O/H)$_{R}$ -- N$_{2}$ diagram.
The long-dashed line shows the excitation parameter $P$ as a function
of the $N_{2}$ for H\,{\sc ii} regions with $0.09 >$ log(O/H)$_{N}$ --
log(O/H)$_{R} > 0.03$, and the short-dashed line for H\,{\sc ii}
regions with $-0.03 >$ log(O/H)$_{N}$ -- log(O/H)$_{R} > -0.09$.
Those H\,{\sc ii} regions with incorrect (O/H)$_{N}$ abundances are
shifted from the O/H $= f$(N$_{2}$) relation in the (O/H)$_{R}$ --
N$_{2}$ diagram.  Panel $c$ shows the same as Panel $a$ but for
spaxels with galactocentric distances from 0.775 to 0.825 $R_{g}$.

Panel $b$ of Fig.~\ref{figure:ln2-p-manga} shows the N$_{2}$/R$_{2}$
line ratio as a function of the $N_{2}$ line intensity for spaxels in
central regions ($R_{g} < 0.1$) of several MaNGA galaxies.  Panel $d$
shows the same as Panel $b$ but for spaxels with galactocentric
distances from 0.775 to 0.825 $R_{g}$.

Examination of Fig.~\ref{figure:ln2-p-manga} shows that the excitation
parameter $P$ and the N$_{2}$/R$_{2}$ line ratio for the spaxels in
central regions of the MaNGA galaxies are close to the values in the
H\,{\sc ii} regions in nearby galaxies with similar (O/H)$_{N}$ and
(O/H)$_{R}$ abundances.  Therefore, the central (O/H)$_{N}$ and
(O/H)$_{R}$ abundances in the MaNGA galaxies are also close to each
other.  At the same time, the excitation parameter $P$ and the
N$_{2}$/R$_{2}$ line ratio for the spaxels with galactocentric
distances from 0.775 to 0.825 $R_{g}$ are located around the trend
described by the H\,{\sc ii} regions in which the (O/H)$_{N}$
abundances are higher than the (O/H)$_{R}$ abundances by 0.03 -- 0.09
dex.  Therefore, the (O/H)$_{N}$ values at the optical radii of the
MaNGA galaxies are systematically higher than the (O/H)$_{R}$
abundances.  This provides strong evidence suggesting that the
discrepancy between the (O/H)$_{N}$ and (O/H)$_{R}$ abundances in the
peripheral regions of the MaNGA galaxies is caused by the fact that
the 1D N-calibration produces incorrect abundances in the
spaxels at the peripheral regions of the MaNGA galaxies because the
excitation $P$ and/or the nitrogen-to-oxygen abundance ratio N/O in
those spaxels differ systematically from those values in the
calibrating points with similar (O/H)$_{N}$ and (O/H)$_{R}$
abundances.
\citet{Ho2015} also noted that the radial change of oxygen abundances derived
with two different 1D calibrations can differ significantly when the ionization
parameters change systematically with radius.

Panel $c$ of Fig.~\ref{figure:ohr-ohx-manga} shows the comparison
between the (O/H)$_{P}$ and (O/H)$_{R}$ abundances for our sample of
MaNGA galaxies.  The circles mark the central abundances; the plus
symbols depict the abundances at the optical radii $R_{25}$.
Unfortunately, the (O/H)$_{P}$ at the optical radius $R_{25}$ can be
estimated in a few MaNGA galaxies only (see above).  Again, the solid
line shows (O/H)$_{P}$ as a function of (O/H)$_{R}$ abundance for our
sample of H\,{\sc ii} regions in nearby galaxies, and the long-dashed
line shows the oxygen abundances for those H\,{\sc ii} regions
obtained with the lines $R_{2}$ and $N_{2}$ increased by a factor of
1.3.  Panel $c$ of Fig.~\ref{figure:ohr-ohx-manga} shows that the
central (O/H)$_{P}$ and (O/H)$_{R}$ abundances in the MaNGA galaxies
follow the trend described by the H\,{\sc ii} regions in nearby
galaxies. This is additional evidence suggesting that the DIG
contribution to the MaNGA spaxel spectra cannot be large, that is, the
enhancement of the  $R_{2}$ and $N_{2}$ strengths does not exceed a
factor of $\sim$1.3.

The 1D N2 calibration suggested by \citet{Pettini2004},
\begin{equation} 
12+\log({\rm O/H})_{\rm PP04,N2} = 8.90 + 0.57 \times {\rm N2}
\label{equation:ppn} 
,\end{equation}
and the updated version of the calibration relation suggested by 
\citet{Marino2013},
\begin{equation} 
12+\log({\rm O/H})_{\rm Ma13,N2} = 8.743 + 0.462 \times {\rm N2}
\label{equation:man} 
,\end{equation} 
where
\begin{equation} 
{\rm N2}  = \log\left(\frac{[{\rm N~II}]\lambda6584}{{\rm H}\alpha}\right)
\label{equation:n2} 
,\end{equation} 
are widely used for abundance estimations.  Using these relations, we
also estimated the (O/H)$_{\rm  PP04,N2}$ and (O/H)$_{\rm  Ma13,N2}$
abundances in the MaNGA galaxies.

Panel $a$ of Fig.~\ref{figure:ohn-ohman-ohppn} shows a comparison of
the (O/H)$_{N2,Ma13}$ abundances obtained through the N2 calibration
relation of \citet{Marino2013} as a function of the (O/H)$_{N}$
abundances determined through our expression of the N calibration for
our sample of MaNGA galaxies; the circles indicate the central
(intersect) oxygen abundances. The plus symbols indicate abundances
(intersect values) at the optical radii $R_{25}$.  The lower panel
shows the same as the upper panel but for the (O/H)$_{N2,PP04}$
abundances obtained through the N2 calibration relation of
\citet{Pettini2004}.  Inspection of Fig.~\ref{figure:ohn-ohman-ohppn}
reveals that the oxygen abundances estimated using the same abundance
indicator depend significantly on the calibration relations (on the
sample of the calibrating data points).

Thus, the $R$ calibration produces the most reliable oxygen abundances
in the MaNGA galaxies. \\ The $N$ calibration yields reliable oxygen
abundances for spaxels where the excitation $P$ and the
nitrogen-to-oxygen abundance ratio are similar to those values in the
calibrating points located close to the O/H = $f$(N$_{2}$) relation in
the (O/H)$_{R}$ -- N$_{2}$ diagram (this holds for the central parts
of the MaNGA galaxies), but produces incorrect abundances for spaxels
where the excitation $P$ and/or the nitrogen-to-oxygen abundance ratio
differs from the those values in the calibrating points (in the outer
parts of the MaNGA galaxies). \\ The $S$ calibration results in
underestimated oxygen abundances in the MaNGA galaxies since the
sulphur line S$_{2}$ is increased in the MaNGA spectra by a factor of
up to $\sim 2$ as compared to its counterpart in the H\,{\sc ii}
regions in nearby galaxies of a given value of N$_{2}$.  

The central oxygen abundance and the radial abundance gradient
obtained for each target galaxy using the $R$ calibration are reported
in Table~\ref{table:sample}.

\section{Discussion}

\begin{figure}
\begin{center}
\resizebox{0.95\hsize}{!}{\includegraphics[angle=000]{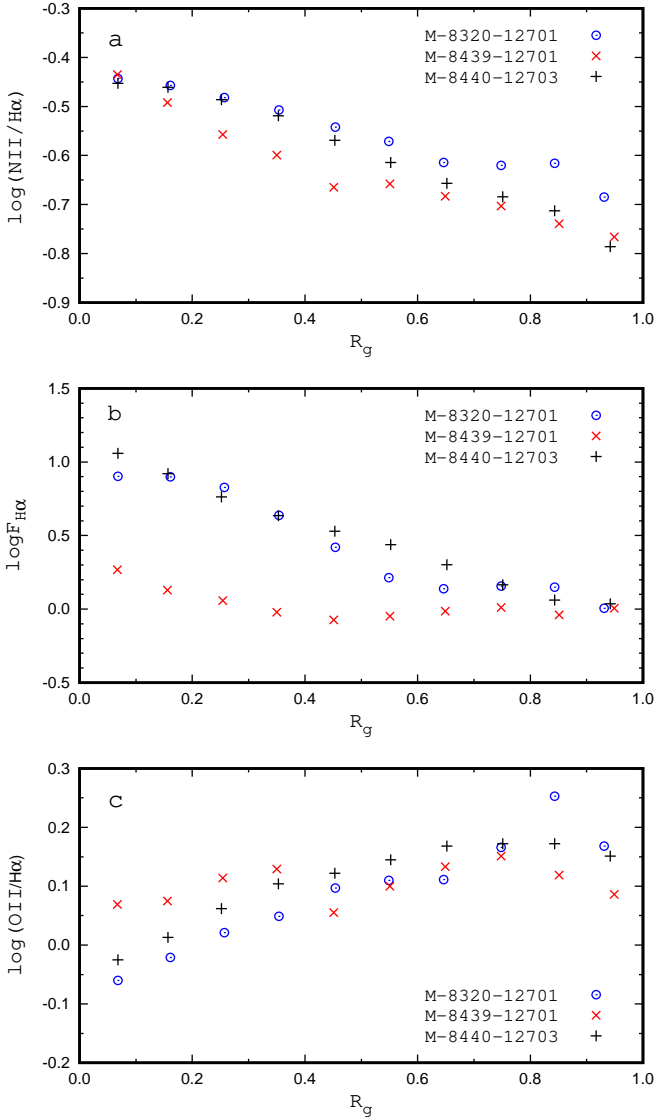}}
\caption{
The  [N\,{\sc ii}]/H$\alpha$ line ratio (Panel $a$), the H$\alpha$
flux (Panel $b$), and the [O\,{\sc ii}]/H$\alpha$ line ratio (Panel
$c$) as a function of galactocentric distance in the discs of three
MaNGA galaxies.  The mean values in bins of 0.1$R_{g}$ are presented. 
}
\label{figure:rg-lfha}
\end{center}
\end{figure}

\begin{figure}
\begin{center}
\resizebox{0.95\hsize}{!}{\includegraphics[angle=000]{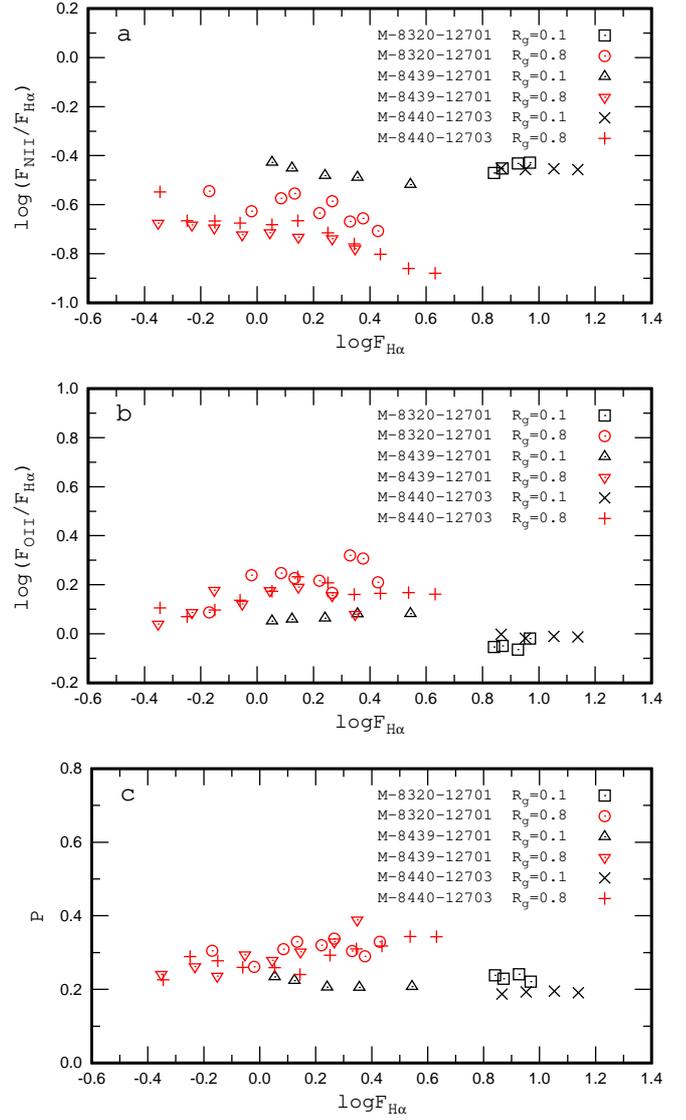}}
\caption{
The [N\,{\sc ii}]/H$\alpha$ (Panel $a$), [O\,{\sc ii}]/H$\alpha$
(Panel $b$) line ratios, and the excitation parameter $P$ (Panel $c$)
as a function of H$\alpha$ flux for spaxels with galactocentric
distances $0.85 > R_{g} > 0.75$ and $0.15 > R_{g} > 0.05$ for
a number of the MaNGA galaxies.  The symbols are the mean values in
bins of 0.1 dex in the H$\alpha$ flux.
}
\label{figure:lfha-ln2ha-rg08}
\end{center}
\end{figure}

\begin{figure*}
\begin{center}
\resizebox{0.95\hsize}{!}{\includegraphics[angle=000]{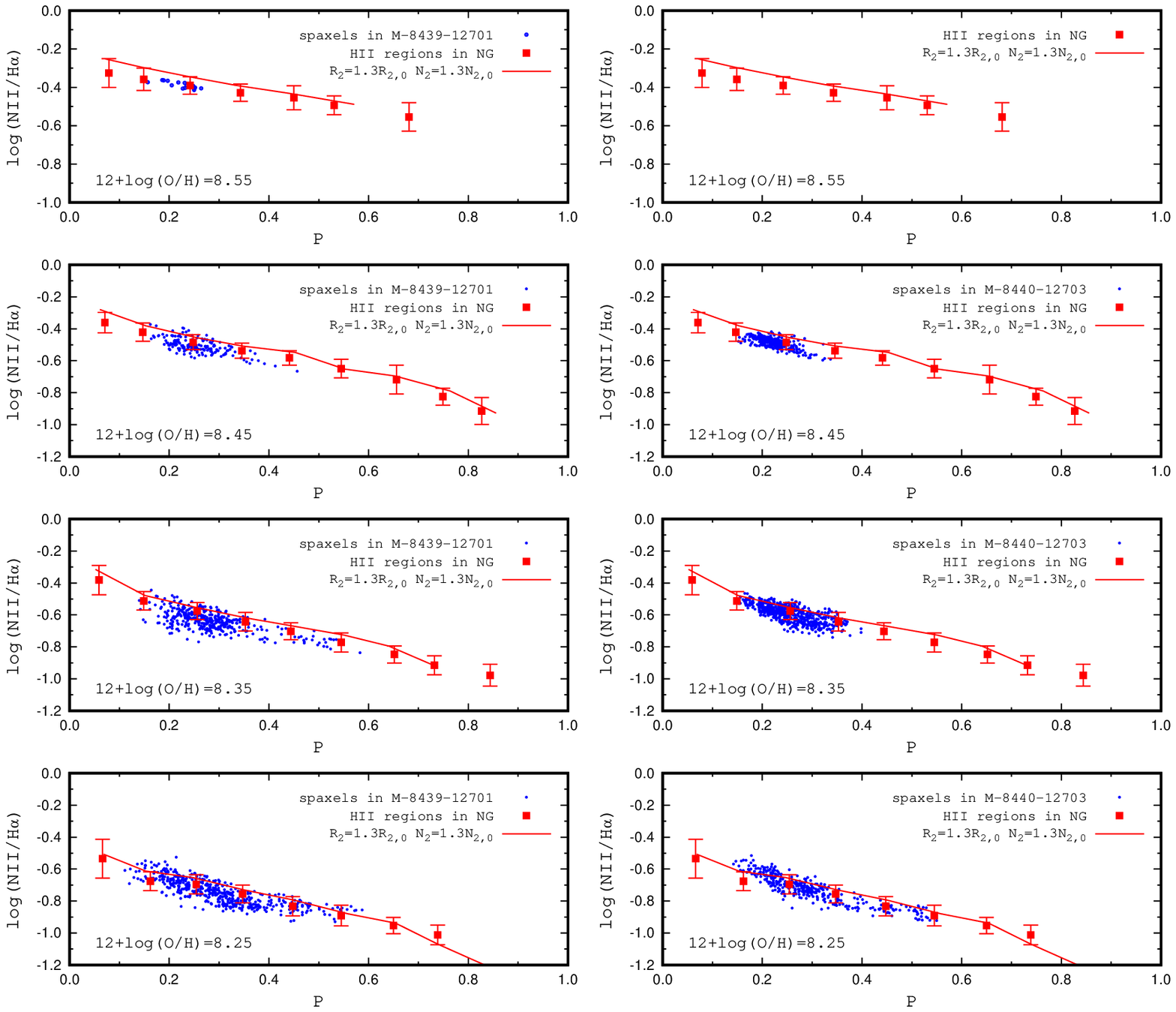}}
\caption{
The left column panels show [N\,{\sc ii}]/H$\alpha$ as a function of
excitation parameter $P$ for the spaxels in the MaNGA galaxy
M-8439-12701 and for the  H\,{\sc ii} regions in nearby galaxies.  The
blue points are spaxels with abundances 12+log(O/H) $\pm 0.05$ in
M-8439-12701 (the value of 12+log(O/H) is indicated in each panel).
The red squares are the H\,{\sc ii} regions in nearby galaxies with
abundances in the same interval.  The squares represent the mean
values of the [N\,{\sc ii}]/H$\alpha$ in bins of 0.05 dex in $P$.  The
bars denote the mean values of the scatter of the [N\,{\sc
ii}]/H$\alpha$ in the bins.  The solid line is for the same H\,{\sc
ii} regions but both the  R$_{2}$ and  N$_{2}$ lines are increased by
a factor of 1.3.  The right column panels show the same for the MaNGA
galaxy M-8440-12703. 
}
\label{figure:p-ln2ha-oh}
\end{center}
\end{figure*}

\citet{Zhang2017} report that the strength of the low-ionization
lines [N\,{\sc ii}], [O\,{\sc ii}], and [S\,{\sc ii}] relative to the
Balmer lines increases with decreasing H$\alpha$ surface brightness
(H$\alpha$ fluxes in spaxels) in the MaNGA galaxies at a fixed
galactocentric distance.  At the same time, the [O\,{\sc iii}]
intensity does not change with the H$\alpha$ surface brightness.
\citet{Zhang2017} concluded that the radiation of the diffuse ionized
gas is dominant in the spectra of spaxels with low H$\alpha$ fluxes.
Thus, on the one hand, the DIG contribution to the spaxel spectra may
be responsible for the difference between the properties (e.g.
excitation) of the spectra of H\,{\sc ii} regions in nearby galaxies
and the spaxel spectra of MaNGA galaxies.  On the other hand, the
excitations of the SDSS fibre spectra are close to the excitation of
the spectra of H\,{\sc ii} regions in nearby galaxies while the
excitation of the MaNGA spaxel spectra differ from that in the H\,{\sc
ii} regions and SDSS fibre spectra (Panel $d$ of
Fig.~\ref{figure:ln2-lx}).  Here we examine whether or not the DIG
contribution is responsible for the difference between the properties
of the spectra of H\,{\sc ii} regions in nearby galaxies and the
spaxel spectra of MaNGA galaxies.

One might expect from the general consideration that the [N\,{\sc
ii}]/H$\alpha$ in the spaxel spectrum depends on the nitrogen
abundance (or on the oxygen abundance and the nitrogen-to-oxygen
abundance ratio) and on the excitation parameter $P$.  Does the global
change of [N\,{\sc ii}]/H$\alpha$ with galactocentric distance in a
galaxy correlate with the change of the H$\alpha$ surface brightness?
Panel $a$ of Fig.~\ref{figure:rg-lfha} shows the  [N\,{\sc
ii}]/H$\alpha$  as a function of galactocentric distance in the discs
of three MaNGA galaxies.  Panel $b$ of Fig.~\ref{figure:rg-lfha} shows
the H$\alpha$ flux $F_{\rm H\alpha}$ in individual spaxels as a
function of galactocentric distance in the discs of those galaxies.
The mean values in bins of $0.1 R_{g}$ are presented.  Since only a
relative comparison of the H$\alpha$ fluxes across the disc of a
galaxy is performed, the measured (and dereddened) H$\alpha$ fluxes in
units of 10$^{-17}$ erg/s/cm$^{2}$/\AA/spaxel
can be used instead of the fluxes reduced to the galaxy's distance.
Because the aperture is the same for all the spaxels, the H$\alpha$
flux in spaxels can be used as the H$\alpha$ surface brightness (in
arbitrary units).   Figure~\ref{figure:rg-lfha} shows that
the changes of the  [N\,{\sc ii}]/H$\alpha$ ratio along the radius are
rather similar in the galaxies studied, while the changes of H$\alpha$ fluxes
are significantly different.  This suggests that [N\,{\sc
ii}]/H$\alpha$ depends more strongly on other parameters than on the
H$\alpha$ flux.

Panel $c$ of Fig.~\ref{figure:rg-lfha} shows the  [O\,{\sc
ii}]/H$\alpha$ ratio as a function of galactocentric distance in the
discs of three MaNGA galaxies.  The change of [O\,{\sc ii}]/H$\alpha$
with galactocentric distance shows the opposite trend in comparison to
the change of [O\,{\sc ii}]/H$\alpha$, that is, the  [O\,{\sc
ii}]/H$\alpha$ increases with galactocentric distance while the
[N\,{\sc ii}]/H$\alpha$ decreases.  Inspection of
Fig.~\ref{figure:rg-lfha} suggests again that the [O\,{\sc
ii}]/H$\alpha$ ratio not only depends on the  H$\alpha$ flux but
also on other parameters.

Figure \ref{figure:lfha-ln2ha-rg08} shows the [N\,{\sc ii}]/H$\alpha$
(Panel $a$), [O\,{\sc ii}]/H$\alpha$ (Panel $b$) line ratios, and the
excitation parameter $P$ (Panel $c$) as a function of H$\alpha$ flux
for spaxels with galactocentric distances $0.85 > R_{g} > 0.75$ and
$0.15 > R_{g} > 0.05$ for a number of the MaNGA galaxies.  The symbols
denote the mean values in bins of 0.1 dex in H$\alpha$ flux.  The
changes of the [N\,{\sc ii}]/H$\alpha$ and [O\,{\sc ii}]/H$\alpha$
ratios in the spaxels of the central parts of the galaxies ($0.15 >
R_{g} > 0.05$) is small.  The changes of the [N\,{\sc ii}]/H$\alpha$
and [O\,{\sc ii}]/H$\alpha$ ratios in the spaxels in the outer parts
of the galaxies ($0.85 > R_{g} > 0.75$) is more appreciable.  Again,
the change of [O\,{\sc ii}]/H$\alpha$ with H$\alpha$ flux shows the
opposite trend than the change of [O\,{\sc ii}]/H$\alpha$, that is, the
[O\,{\sc ii}]/H$\alpha$ ratio increases with H$\alpha$ flux while the
[N\,{\sc ii}]/H$\alpha$ ratio decreases.  Since the excitation $P$
also changes with the H$\alpha$ flux (Panel $c$ of
Fig.~\ref{figure:lfha-ln2ha-rg08}), this may be responsible for the
changes of the strengths of the [N\,{\sc ii}]/H$\alpha$ and [O\,{\sc
ii}]/H$\alpha$ ratios along with the changes of the H$\alpha$ flux.

The panels in the left column of Fig.~\ref{figure:p-ln2ha-oh} show the
[N\,{\sc ii}]/H$\alpha$ ratio as a function of the excitation
parameter $P$ for the spaxels in the MaNGA galaxy M-8439-12701 and for
the  H\,{\sc ii} regions in nearby galaxies with abundances
12+log(O/H) $\pm$ 0.05 (the value of 12+log(O/H) is indicated in each
panel).  The blue points are spaxels in M-8439-12701 and the red
squares are the H\,{\sc ii} regions in nearby galaxies. The squares
represent the mean values of the [N\,{\sc ii}]/H$\alpha$ ratio in bins
of 0.05 dex in $P$.  The bars denote the mean values of the scatter of
[N\,{\sc ii}]/H$\alpha$ in the bins.  The solid line represents the
same H\,{\sc ii} regions in nearby galaxies but both the R$_{2}$ and
N$_{2}$ lines are increased by a factor of 1.3.  Inspection of
Fig.~\ref{figure:p-ln2ha-oh} reveals that the [N\,{\sc ii}]/H$\alpha$ -- $P$
diagrams for the MaNGA spaxel spectra are compatible with the one for
the spectra of H\,{\sc ii} regions in nearby galaxies.  Differences
are seen only in low-metallicity, low-excitation regions. 

Thus, there is little room for a significant increase of the strength
of the low-ionization lines, [O\,{\sc ii}] and [N\,{\sc ii}], relative
to the Balmer lines in the MaNGA spaxel spectra due to the
contribution of the DIG radiation.  This suggests that the demarcation
line between AGNs and H\,{\sc ii} regions in the BPT diagram defined
by \citet{Kauffmann2003} is a good criterium to reject the spectra
with a significantly distorted strength of the [N\,{\sc ii}] and
[O\,{\sc ii}] lines.

\section{Summary}

The publicly available data obtained by the MaNGA survey form the basis
of our current study. We measured the emission lines in the spectrum
of each spaxel from the MaNGA datacubes for 47 late-type galaxies.
The surface brightness in the SDSS $g$ and $r$ bands was obtained from
broadband SDSS images created from the data cube. 

It has been suggested that the strength of the low-ionization lines,
R$_{2}$, N$_{2}$, and S$_{2,}$ can be increased (relative to the Balmer
lines) in (some) spaxel spectra of the MaNGA survey due to a
contribution of the radiation of the diffuse ionized gas and that,
consequently, the abundances derived from the spaxel spectra through
strong-line methods may have large errors.  We examine this
expectation by comparing the behaviour of the line intensities and the
abundances estimated through different calibrations for samples of
slit spectra of H\,{\sc ii} regions in nearby galaxies, of the fibre
spectra from the SDSS, and of the spaxel spectra of the MaNGA survey.
We find that the mean value of the S$_{2}$ strength in the SDSS fibre
and MaNGA spaxel spectra is increased significantly (up to a factor of
$\sim$2) depending on the metallicity.  The mean distortion of R$_{2}$
and N$_{2}$ is less than a factor of $\sim$1.3.  This suggests that
the demarcation line between AGNs and H\,{\sc ii} regions in the BPT
diagnostic diagram defined by \citet{Kauffmann2003} is a relatively
reliable criterium to reject the spectra with significantly distorted
strengths of the [N\,{\sc ii}] and [O\,{\sc ii}] lines. 

We determined and compared the oxygen abundances in the SDSS galaxies
through different calibrations based on two catalogues of emission
line measurements aiming to estimate the uncertainty in the oxygen
abundances due to the uncertainties in the data reduction and line
flux measurements. We find that the uncertainty in the abundances are
small ($\la 0.02$ dex) for objects with metallicity from 12+log(O/H)
$\sim 8.1$ to 12 + log(O/H) $\sim 8.55$ and are higher for objects
outside this interval (with lower and higher metallicities).

We analysed the abundances estimated through different calibrations
for samples of the slit spectra of the H\,{\sc ii} regions in nearby
galaxies, of the fibre spectra of the SDSS, and of the spaxel spectra
of the MaNGA survey, aiming to establish which calibration produces the
most reliable abundances from the MaNGA spaxel spectra. \\ We find
that the 3D $R$ calibration using the R$_{2}$ and
N$_{2}$ lines produces the most reliable oxygen abundances in the
MaNGA galaxies; and that the 1D $N$ calibration produces
reliable oxygen abundances for spaxels where the excitation $P$ and
the nitrogen-to-oxygen abundance ratio are close to those values in
the calibrating points located close to the O/H = $f$(N$_{2}$)
relation in the  (O/H)$_{R}$ -- N$_{2}$ diagram   (i.e. in
calibrating H\,{\sc ii} regions with similar (O/H)$_{N}$ and
(O/H)$_{R}$ abundances), but produces incorrect abundances for spaxels
where the excitation $P$ and/or the nitrogen-to-oxygen abundance ratio
differ from the values in the calibrating points. 

We determined the abundance distributions (and surface brightness
profiles) within the optical radii of the discs of 47 galaxies based
on the spaxel spectra from the MaNGA survey.  Those data will be used
in a forthcoming paper.

\section*{Acknowledgements}

We are grateful to the referee for his/her constructive comments. \\
L.S.P., E.K.G., and I.A.Z.\  acknowledge support within the framework
of Sonderforschungsbereich (SFB 881) on ``The Milky Way System''
(especially subproject A5), which is funded by the German Research
Foundation (DFG). \\ 
L.S.P.\ and I.A.Z.\ are grateful for the hospitality of the
Astronomisches Rechen-Institut at Heidelberg University, where part of
this investigation was carried out. \\
I.A.Z.\ acknowledges the support of the Volkswagen Foundation 
under the Trilateral Partnerships grant No.\ 90411. \\
This work was partly funded by the subsidy allocated to Kazan Federal 
University for the state assignment in the sphere of scientific 
activities (L.S.P.).  \\ 
We acknowledge the use of the HyperLeda database (http://leda.univ-lyon1.fr). \\
This research made
use of Montage, funded by the National Aeronautics and Space
Administration's Earth Science Technology Office, Computational
Technnologies Project, under Cooperative Agreement Number NCC5-626
between NASA and the California Institute of Technology. The code is
maintained by the NASA/IPAC Infrared Science Archive. \\ 
Funding for SDSS-III has been provided by the Alfred P. Sloan Foundation,
the Participating Institutions, the National Science Foundation,
and the U.S. Department of Energy Office of Science.
The SDSS-III web site is http://www.sdss3.org/. \\
SDSS-III is managed by the Astrophysical Research Consortium
for the Participating Institutions of the SDSS-III Collaboration
including the University of Arizona, the Brazilian Participation Group,
Brookhaven National Laboratory, Carnegie Mellon University,
University of Florida, the French Participation Group,
the German Participation Group, Harvard University,
the Instituto de Astrofisica de Canarias,
the Michigan State/Notre Dame/JINA Participation Group,
Johns Hopkins University, Lawrence Berkeley National Laboratory,
Max Planck Institute for Astrophysics,
Max Planck Institute for Extraterrestrial Physics,
New Mexico State University, New York University,
Ohio State University, Pennsylvania State University,
University of Portsmouth, Princeton University,
the Spanish Participation Group, University of Tokyo,
University of Utah, Vanderbilt University, University of Virginia,
University of Washington, and Yale University. \\ 
Funding for the Sloan Digital Sky Survey IV has been provided by the
Alfred P. Sloan Foundation, the U.S. Department of Energy Office of Science,
and the Participating Institutions. SDSS-IV acknowledges
support and resources from the Center for High-Performance Computing at
the University of Utah. The SDSS web site is www.sdss.org. \\
SDSS-IV is managed by the Astrophysical Research Consortium for the 
Participating Institutions of the SDSS Collaboration including the 
Brazilian Participation Group, the Carnegie Institution for Science, 
Carnegie Mellon University, the Chilean Participation Group,
the French Participation Group, Harvard-Smithsonian Center for Astrophysics, 
Instituto de Astrof\'isica de Canarias, The Johns Hopkins University, 
Kavli Institute for the Physics and Mathematics of the Universe (IPMU) / 
University of Tokyo, Lawrence Berkeley National Laboratory, 
Leibniz Institut f\"ur Astrophysik Potsdam (AIP),  
Max-Planck-Institut f\"ur Astronomie (MPIA Heidelberg), 
Max-Planck-Institut f\"ur Astrophysik (MPA Garching), 
Max-Planck-Institut f\"ur Extraterrestrische Physik (MPE), 
National Astronomical Observatories of China, New Mexico State University, 
New York University, University of Notre Dame, 
Observat\'ario Nacional / MCTI, The Ohio State University, 
Pennsylvania State University, Shanghai Astronomical Observatory, 
United Kingdom Participation Group,
Universidad Nacional Aut\'onoma de M\'exico, University of Arizona, 
University of Colorado Boulder, University of Oxford, University of Portsmouth, 
University of Utah, University of Virginia, University of Washington, University of Wisconsin, 
Vanderbilt University, and Yale University.

\end{document}